\documentclass[a4paper]{article}
\usepackage{a4wide}
\usepackage[UKenglish]{babel}
\usepackage{authblk}
\usepackage{amsmath,amsthm,amssymb}
\usepackage{accents}
\usepackage{graphicx,pdfpages}
\graphicspath{{./}{figs/}{other/}}
\usepackage{subcaption}
\usepackage{booktabs}
\usepackage[ruled,vlined,linesnumbered]{algorithm2e}
\usepackage[inline]{enumitem}
\usepackage[numbers,square]{natbib}

\usepackage[colorlinks=true,linkcolor=blue,citecolor=red]{hyperref}

\newcommand{\N}{\mathbb{N}}
\newcommand{\R}{\mathbb{R}}

\DeclareMathOperator\Lognormal{Lognormal}
\DeclareMathOperator\BiLognormal{BiLognormal}
\DeclareMathOperator\Bernoulli{Bernoulli}
\DeclareMathOperator\IC{IC}
\DeclareMathOperator\CCDF{\mathcal R}

\newtheorem{dfn}{Definition}
\newtheorem{rmk}[dfn]{Remark}
\newtheorem{hyp}[dfn]{Assumption}

\newsavebox{\tikzextfigbox}
\providecommand\newcaption[2]{%
  \captionsetup{width=#2}%
  \caption{#1}%
}
\newcommand{\tikzextfig}[1]{%
  \sbox{\tikzextfigbox}{\includegraphics{#1}}%
  \ifdim\wd\tikzextfigbox>\linewidth
    \resizebox{\linewidth}{!}{\usebox{\tikzextfigbox}}%
  \else
    \usebox{\tikzextfigbox}%
  \fi
}

\allowdisplaybreaks

\title{Modelling population distribution across cities on spatial networks using the kinetic theory of multi-agent systems}
\author{Valerio Taralli}
\author{Andrea Tosin}
\affil{{\small Department of Mathematical Sciences ``G. L. Lagrange'', Politecnico di Torino, Italy}}
\date{}
		
\begin{document}
\maketitle

\begin{abstract}
We investigate the urban population distribution of the Italian region of Sardinia through the kinetic theory of multi-agent systems on spatial networks, interpreting cities both as agents and vertices of a graph, and representing interurban links as edges. We consider various emigration rules to model the interactions between agents, obtaining two main results: the bimodal lognormal fit of the total population distribution among cities and the Pareto fit of the tail. Overall, our model provides results consistent with the actual distribution derived from ISTAT data of Sardinia; furthermore, the bimodal lognormal fit generally follows the distribution tail more accurately than the Pareto one, although the two coincide in the first half.

\medskip


\medskip

\end{abstract}

\section{Introduction}
\label{secIntroduction}
Population distribution across cities has been sporadically analysed in the past. The first was Auerbach \cite{Auerbach1913} who, at the beginning of the 20th century, noted a feature formalised by Zipf\footnote{Actually, as Zipf himself admits in~\cite{Zipf2016}, he was not the first to discover such a law.}~\cite{Zipf2016} almost fifty years later, albeit initially in a linguistic context. Zipf's empirical law states the following relationship between the frequency $\nu$ of a word in a language and its rank $r$ within the most frequent words of that language:
\begin{equation}
    \nu\propto\frac{1}{r}.
    \label{eqZipfsLaw}
\end{equation}
To give a concrete example, this means that there exists a constant $C>0$ such that if the tenth most frequent word appears $2000$ times, then $2000\approx C/10$. The same law described in~\eqref{eqZipfsLaw} can be found in many other contexts, including the distribution of population across cities if one replaces $\nu$ with the \textit{population size} $s$~\cite[Fig. 9-2]{Zipf2016}. However, this is true only for $s\gg 1$, namely in the tail of the distribution, contrary to Zipf's original linguistic context, where~\eqref{eqZipfsLaw} is instead expected to hold always.
        
More generally, Zipf's law can be described by a Pareto distribution: let $S\in\R_+:=[0,\,+\infty)$ be the random variable describing a city size and let $f_S=f_S(s)$ be its probability distribution, which allows one to interpret $f_S(s)\,ds$ as the fraction of cities with $S\in [s,\,s+ds]$. Then, $S$ is said to follow a Pareto distribution in the tail if
\begin{equation}
    \int_{s}^{+\infty}f_S(z)\,dz \approx \frac{c}{s^\beta}, \quad s\gg 1,
    \label{eqComplementaryCumulativePareto}
\end{equation}
where $\beta,\,c\in\R_+$ are parameters. In particular, the exponent $\beta$ is called the Pareto index. For $\beta=1$,~\eqref{eqComplementaryCumulativePareto} is analogous to~\eqref{eqZipfsLaw} with $s$ playing the role of $r$.
        
More recently, it has been discovered~\cite{Eeckhout2004,Gualandi2019HBLD} that, apart from the tail, $f_S$ is well fitted by the probability density function of a lognormal random variable $X\sim\Lognormal(\mu,\,\sigma)$, specifically
\begin{equation}
    f_X(x)=\frac1{x\sigma\sqrt{2\pi}}\exp\!\left(-\frac{(\log{x}-\mu)^2}{2\sigma^2}\right).
    \label{eqLognormalDistribution}
\end{equation}
As the name suggests, the peculiarity of this distribution is that $\log{X}\sim\mathcal{N}(\mu,\,\sigma)$, where $\mathcal{N}(\mu,\,\sigma)$ denotes the Gaussian distribution with mean $\mu$ and standard deviation $\sigma$, cf. Figure~\ref{figLognormalFunctionExample}.
        
Furthermore, Gualandi and coauthors~\cite{Gualandi2019HBLD,Gualandi2019SDC} realised that the entire distribution is well fitted by a bimodal lognormal (here also referred to as bilognormal), namely a convex combination of~\eqref{eqLognormalDistribution}:
\begin{equation}
    f_S(s)\approx \zeta f_{X_1}(s)+(1-\zeta)f_{X_2}(s),
    \label{eqBimodalLognormalDistribution}
\end{equation}
with $\zeta\in(0,\,1)$, $X_i\sim\Lognormal(\mu_i,\,\sigma_i)$ for $i=1,\,2$, whence $S\approx\BiLognormal(\zeta,\,\mu_1,\,\sigma_1,\,\mu_2,\,\sigma_2)$.
        
Building on these preliminary observations, in this paper we aim to model, by means of the Kinetic Theory of Multi-Agent Systems (KTMAS), the interactions between cities connected on a spatial graph, thereby attempting to reproduce a population distribution with characteristics analogous to those actually observed in the Italian region of Sardinia -- mainly bilognormal body and Pareto tail. Upon achieving such a goal, we interpret and explain the migration phenomenon through the emigration laws at the basis of the interactions between cities in our kinetic model.
        
In the present work, we regard cities as agents characterised by their population size $s\in\R_+$, and assume that they interact, i.e. they exchange individuals, through an underlying spatial graph structure. Except for the network, such a description is analogous to the classical one of gas molecules characterised by their position and velocity in the kinetic theory of gases. As a matter of fact, in the literature, at least to the authors' knowledge, there are no contributions dealing simultaneously with KTMAS, population distribution across cities, and graphs, despite the natural modelling representation that these elements provide. More precisely:
\begin{itemize}
\item Some authors either apply the KTMAS to cities without considering any underlying graph topology~\cite{Gualandi2019SDC} or, conversely, consider the graph structure without applying it~\cite{Berliant2018}.
\item Other authors view graph vertices as places where agents live and interact, thus creating a large system of coupled Boltzmann-type equations~\cite{Loy2021}.
\item The contribution most closely related to the present paper applies the aforementioned theory to the context of social networks~\cite{Nurisso2024}.
\end{itemize}

This gap is probably due to the fact that developments in graph theory applied to the KTMAS have been quite recent and primarily focused on the vertex-place perspective rather than the vertex-agent one. Therefore, the primary contribution of this work is showing that the second perspective is as valid as the first, despite being less common. Therefore, this paper presents a formal application-oriented exploration of the vertex-agent perspective; for preliminary analytical investigations, we refer the interested reader to~\cite{taricco2026PREPRINT}.

Overall, the paper is organised as follows. The second section summarises the essential notions of the graph-mediated KTMAS and contextualises them within the problem of population distribution across cities. The third section presents numerical results for various emigration rules applied to the Italian region of Sardinia. Finally, the fourth section draws conclusions and outlines potential directions for future research.

\section{Outline of the KTMAS on graphs}
\label{secNotesKTMAS}
In this section, we briefly discuss some assumptions on the type of graph used and the two Boltzmann-type equations on graphs that we use to formalise our mathematical model. For the general principles of KTMAS, we refer the interested reader to~\cite{Loy2025EKTMAS,PareschiIMS2013}.

\subsection{Networks and cities}
\label{secNetworksAndCities}
We begin by recalling the definition of graph:
            
\begin{dfn} \label{defGraph}
A \emph{graph} is a set $\mathcal G:=(\mathcal I,\,\mathcal E)$, where $\mathcal I\subseteq\N_+$ is the set of vertex indices and $\mathcal E$ is the set of edges, which are pairs of indices in $\mathcal I$. Two vertices $i,\,j\in\mathcal I$ are connected if and only if $(i,\,j)\in\mathcal E\subseteq\N_+\times\N_+$.
\end{dfn}
            
Next, we discuss how to interpret a link between two vertices, viz. cities, given the multitude of possible urban connections~\cite{Barthelemy2011}. In fact, a city can be represented at many levels by means of a graph:
\begin{enumerate*}[label=(\roman*)]
\item \textit{intraurban} level, representing either streets as edges and their intersections as vertices or transport networks (tram, bus, and metro);
\item \textit{interurban} level, taking individually into consideration various transport networks (for instance, railway, naval, air).
\end{enumerate*}
Recalling the introductory section, it is clear that in order to predict the population distribution across cities the following observations hold: on the one hand, intraurban representations should be discarded, because in that context people's movements take place within a single city; on the other hand, interurban representations should be considered simultaneously and not individually, since each individual may choose different transportation modes to move.
    
This is why the graph we refer to in our model is the one related to commuting movements across cities~\cite{Barthelemy2011,DeMontis2007}, which holistically shows all their possible connections regardless of the chosen transportation mode. Moreover, it shows realistic connections associated with daily rather than extraordinary movements, such as e.g., holidays or medical visits. Upon choosing such a representation, we need to establish the type of graph we deal with:
\begin{hyp} \label{hypGraphType}
The graph $\mathcal G$ is \emph{undirected}, hence its adjacency matrix is symmetric.
\end{hyp}

A comment on Assumption~\ref{hypGraphType} is in order. If one city can interact with another, then the reverse is also possible, as the two cities are connected. Thus, the underlying graph is undirected, or equivalently, its connectivity structure is symmetric. This makes Assumption~\ref{hypGraphType} a natural choice in the context of our problem. However, one city may interact, viz. exchange population, with another without the reverse interaction necessarily taking place. Hence, interactions between cities need not be symmetric, even when the underlying graph is undirected. The symmetry of the graph and that of the interactions are therefore two independent properties.

In addition to this, we are only interested in predicting the distribution of the population with respect to an \textit{a priori} fixed graph topology, which leads to the following further assumption:
\begin{hyp} \label{hypStaticGraph}
The graph $\mathcal G$ is \emph{non-coevolving}, i.e. the sets $\mathcal I$ and $\mathcal E$ do not vary in time.
\end{hyp}
Clearly, this is a substantial simplification, since, in reality, the graph topology, particularly the interurban connections in $\mathcal E$, co-evolves following the development of the cities themselves.
            
Finally, we observe that this type of graph is spatial~\cite{Barthelemy2011,DeMontis2007}, meaning that its vertices are points in an Euclidean space, even though, for the moment, this is a somewhat formal characteristic. However, it will be useful to keep it in mind when interpreting certain emigration rules in Section~\ref{secInterpretations}. Moreover, contrary to what one might think at first glance, our graph is not scale-free \cite{Barabasi1999}, precisely because of its spatial nature~\cite{Barthelemy2011,Broido2019}. This, however, does not rule out the existence of vertices that are more central than others, but only implies that the maximum degree of a vertex is bounded above due to physical constraints.
    
\subsection{Boltzmann-type equation on the exact graph topology}
\label{secExactBoltzmannType}
The KTMAS is centred around agents interacting with one another, thereby changing their microscopic state. However, usually the precise knowledge of the microscopic state of a single agent is neither useful nor necessary, which is why a statistical approach is taken by regarding microscopic states as time-evolving random variables forming a stochastic process. To derive an evolution equation for the probability distribution, say $f$, of the microscopic state of a representative agent, four assumptions are formulated, cf.~\cite{Loy2025EKTMAS}:
\begin{hyp} \label{hypBoltzmannType}
\begin{enumerate}[before=\leavevmode,label=(\roman*)]
\item\label{hypBoltzmannTypeHomogeneousDensity} The probability distribution $f$ is homogeneous in space.
\item\label{hypBoltzmannTypeBinaryInteractions} Non-pairwise interactions between agents are neglected.
\item\label{hypBoltzmannTypeIndistinguishableAgents} Agents are indistinguishable, and each of them represents the ensemble.
\item\label{hypBoltzmannTypeMolecularChaos} Two interacting agents are statistically independent at the moment of the interaction.
\end{enumerate}
\end{hyp}

Apparently, an inconsistency arises from Assumption~\ref{hypBoltzmannType}\ref{hypBoltzmannTypeIndistinguishableAgents} if agents are confused with graph vertices: the former are indistinguishable, but the underlying network makes the latter distinct due to the indices in $\mathcal I$. To bypass such a difficulty of the theory, it is sufficient to include the vertex index in the agent's microscopic state, considering it as a random variable. As a result, the microscopic state of a generic agent of the system is the pair $(I,\,S_t)$, where $S_t\in\R_+$ is the random variable expressing the population size at time $t\geq 0$ of a representative agent, viz. city, of the system. Due to~Assumption~\ref{hypStaticGraph}, the random variable $I$ is constant in time.

\begin{rmk} \label{rmkPopulationNumericalNature}
It is natural that fractions of people do not exist, hence in principle $S_t\in\mathbb N$. Nevertheless, the most realistic choice is not always the easiest modelling-wise. At any rate, if necessary, it is possible to gauge the actual population size by rounding it to the nearest natural number.
\end{rmk}

\begin{figure}[!t]
  \centering
  \tikzextfig{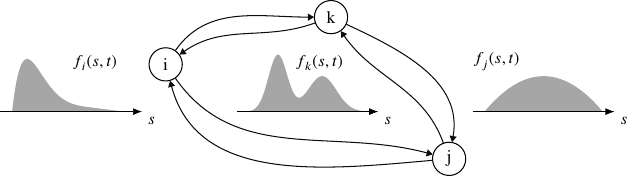}
  \newcaption{Graphical representation of the KTMAS on graphs}{\linewidth}
  \label{figGraphicRepresentationKTMASNetworks}
\end{figure}

Reasoning in this way, we describe statistically the microscopic state $(I,\,S_t)$ of a representative city with the joint probability distribution $f:\mathcal I\times\R_+\times\R_+\to\R_+$, which is discrete in the first variable $i\in\mathcal I$ and continuous in the second variable $s\in\R_+$, and takes therefore the form
\begin{equation}
    f(i,s,t):=\frac{1}{N}\sum_{j\in\mathcal I}f_j(s,t)\otimes\delta(i-j),
    \label{eqUrbanJointDensity}
\end{equation}
where $N:=\vert\mathcal I\vert$ is the total number of cities in the graph, while $\delta(\cdot)$ denotes the Dirac delta centred at the origin. On the other hand, $f_i:\R_+\times\R_+\to\R_+$ is the probability distribution of $S_t$ conditioned to the $i$-th city, cf. Figure~\ref{figGraphicRepresentationKTMASNetworks}. This way, $f$ integrates to $1$ over its domain:
$$ \int_{\mathcal I}\int_{\R_+}f(i,s,t)\,ds\,di=\frac{1}{N}\sum_{i\in\mathcal I}\int_{\R_+}f_i(s,t)\,ds=1, \qquad \forall\,t\geq 0. $$
            
When two cities interact they exchange population, thereby altering their respective microscopic states. This is modelled by the following general interaction rules:
\begin{equation}
    S'_t=\psi(I,S_t,I_\ast,S_{\ast,t}), \qquad S_{\ast,t}'=\psi_\ast(I,S_t,I_\ast,S_{\ast,t}),
    \label{eqUrbanInteractionRules}
\end{equation}
where the functions $\psi,\,\psi_\ast:\mathcal{I}\times\R_+\times\mathcal{I}\times\R_+\to\R_+$ describe the post-interaction microscopic states. However, two randomly picked cities can interact only if they are connected by an edge of the graph $\mathcal G$. This can be taken into account via the Bernoulli random variable
\begin{equation}
    \Theta\sim\Bernoulli(a_{I,I_\ast}\Delta t), \qquad
        \text{with} \quad
        a_{I,I_\ast}:=
        \begin{cases}
            1 & \text{if } (I,\,I_\ast)\in\mathcal E \\
            0 & \text{otherwise},
        \end{cases}
    \label{eqUrbanInteractionBernoulli}
\end{equation}
where $a_{I,I_\ast}$ is the entry of the adjacency matrix of $\mathcal G$ corresponding to the pair of indices $(I,\,I_\ast)$ and $\Delta{t}>0$ is a discrete time step corresponding to the characteristic duration of an interaction. For consistency, we require $\Delta{t}\leq 1$, so that the parameter of the random variable $\Theta$ is well defined.

Combining~\eqref{eqUrbanInteractionRules} and~\eqref{eqUrbanInteractionBernoulli}, the following agent-based interaction algorithm is obtained:
\begin{equation}
    \begin{cases}
        S_{t+\Delta t}=(1-\Theta)S_t+\Theta S'_t \\
        S_{\ast,t+\Delta t}=(1-\Theta)S_{\ast,t}+\Theta S'_{\ast,t},
    \end{cases}
    \label{eqUrbanInteractionAlgorithm}
\end{equation}
consistently with the fact that, in an urban context, if a city exchanges population with another both change their population size.

Let now $\varphi:\R_+\to\R$ be an arbitrary observable quantity. By averaging~\eqref{eqUrbanInteractionAlgorithm} over all possible observable quantities and taking the continuous-time limit $\Delta{t}\to 0^+$, one obtains, cf.~\cite{Nurisso2024}:
\begin{align}
    \begin{aligned}[b]
        \frac{d}{dt}&\int_{\R_+}\varphi(s)f_i(s,t)\,ds \\
        & =\frac{1}{N}\sum_{i_\ast\in\mathcal I}a_{i,i_\ast}\iint_{\R_+^2}\frac{\varphi(s')+\varphi(s_\ast')-\varphi(s)-\varphi(s_\ast)}{2}
            f_i(s,t)f_{i_\ast}(s_\ast,t)\,ds\,ds_\ast, \qquad \forall\,i\in\mathcal{I},
    \end{aligned}
    \label{eqUrbanAsymmetricHomogeneousWeakFormBoltzmannType}
\end{align}
which is a system of Boltzmann-type equations in weak form for the in-vertex probability distributions $\{f_i\}_{i=1}^{N}$, based on the exact topology of the connections of $\mathcal G$ as encoded in the $N\times N$ adjacency matrix $\mathbf{A}:=(a_{i,i_\ast})_{i,i_\ast=1}^{N}$.

In~\eqref{eqUrbanAsymmetricHomogeneousWeakFormBoltzmannType}, $s',\,s_\ast'$ are the post-interaction population sizes of two interacting cities, which exchange inhabitants on the basis of the interaction rules~\eqref{eqUrbanInteractionRules}. In particular, they are realisations of the random variables $S_t',\,S_{\ast,t}'$.

\subsection{Boltzmann-type equation on an approximated graph topology}
\label{secApproximatedBoltzmannType}
In~\cite{Nurisso2024}, an approximation of~\eqref{eqUrbanAsymmetricHomogeneousWeakFormBoltzmannType} has been proposed based on a corresponding approximation of the adjacency matrix $\mathbf A$. Such an approximation is introduced so as to make it possible to pass to the \textit{dense graph limit} $N\to\infty$, thereby obtaining a single Boltzmann-type equation, which does not require the precise knowledge of the graph connections.

To this purpose, we recall that $\mathbf A$ is assumed to be symmetric, cf. Assumption~\ref{hypGraphType}, and we give the following
\begin{dfn} \label{defDegreeMatrix}
Let $\mathbf{1}:=(1,\,1,\,\dots,\,1)^\top\in\R^N$, and define
$$ \boldsymbol{\xi}:=\mathbf{A}\mathbf{1}, $$
the vector of the \emph{degrees} of the vertices of $\mathcal G$; i.e., $\xi_i:=\sum_{i_\ast\in\mathcal I}a_{i,i_\ast}\in\{0,\,\dots,\,N\}$ is the number of edges incident on vertex $i\in\mathcal I$, for $i=1,\,\dots,\,N$. Define moreover
$$ \mathbf{B}:=\frac{\boldsymbol{\xi}\boldsymbol{\xi}^\top}{d_N}, \qquad \text{with} \quad d_N:=\mathbf{1}^\top\mathbf{A}\mathbf{1}. $$
Then, since $\mathbf{B}\mathbf{1}=\mathbf{A}\mathbf{1}$, $\mathbf{B}\in\R^{N\times N}$ is a rank-one matrix with the same degrees as $\mathbf{A}$.
\end{dfn}

Replacing $\mathbf A$ with $\mathbf B$ corresponds to approximating $\mathcal G$ with a \textit{random graph} with the same degree sequence as $\mathcal G$, see~\cite{bollobas1980EJC,molloy1995RSA}. Using $\mathbf B$ instead of $\mathbf A$ in~\eqref{eqUrbanAsymmetricHomogeneousWeakFormBoltzmannType} and, at the same time, introducing the \textit{normalised} degree variable
$$ x:=\frac{\xi}{N}\in [0,\,1], $$
it can be shown, see~\cite{Nurisso2024,taricco2026PREPRINT}, that~\eqref{eqUrbanAsymmetricHomogeneousWeakFormBoltzmannType} converges, in the limit $N\to\infty$, to the following weak Boltzmann-type equation:
\begin{multline}
    \frac{d}{dt}\int_{\R_+}\varphi(s)g(x,s,t)\,ds \\
    =\int_0^1\iint_{\R_+^2}\frac{x x_\ast}{d}\cdot\frac{\varphi(s')+\varphi(s_\ast')-\varphi(s)-\varphi(s_\ast)}{2}
        g(x,s,t)g(x_\ast,s_\ast,t)\,ds\,ds_\ast\,dx_\ast,
    \label{eqRelaxedGraphBoltzmannType}
\end{multline}
where:
\begin{itemize}
\item $g$ is the limit probability distribution that $f$ in~\eqref{eqUrbanJointDensity} converges to, up to a suitable scaling, in the dense graph limit.
\item $d:=\int_0^1\int_{\R_+}xg(x,s,t)\,ds\,dx$ is the dense-graph-limit analogue of $d_N$ in Definition~\ref{defDegreeMatrix}, see~\cite{Nurisso2024}.
\end{itemize}

It is worth stressing that, in the limit, $x\in [0,\,1]$ plays the role of a (continuous) \textit{latent variable}, which accounts for the topology of the connections among the cities.

\begin{figure}[!t]
  \centering
  \tikzextfig{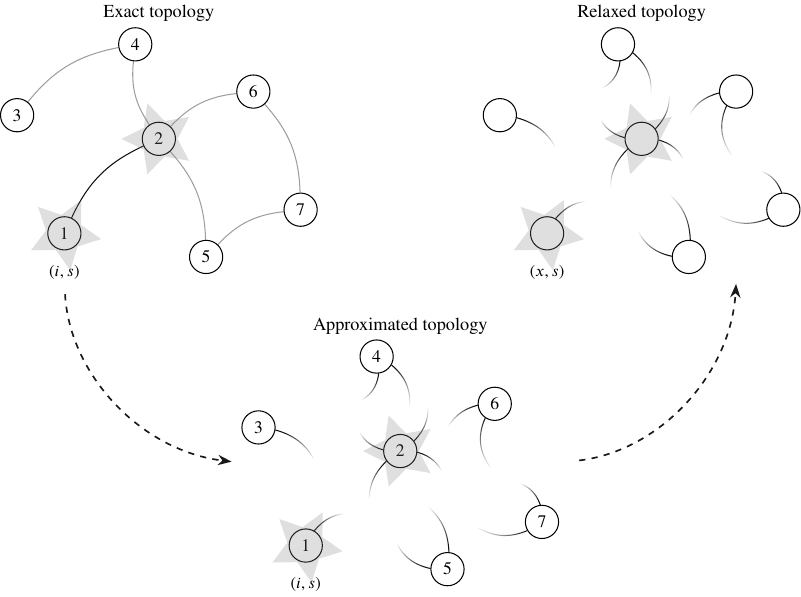}
  \newcaption{Graphical representation of the KTMAS approximation on graphs}{.85\linewidth}
  \label{figGraphicalRepresentationApproximation}
\end{figure}

The implications of~\eqref{eqRelaxedGraphBoltzmannType} are profound: approximating $\mathbf A$ with $\mathbf B$ implies losing the distinguishability of the vertices of the graph, as it is well represented by the passage from the index $i$ to the latent variable $x$. This way, the graph topology is relaxed, cf. Figure~\ref{figGraphicalRepresentationApproximation}; specifically, it remains solely as a degree distribution. Such a reduction is akin to blurring the connection details of the vertices: one transitions from a specific graph to a class of random graphs sharing the same degree distribution. Therefore,~\eqref{eqRelaxedGraphBoltzmannType} is closer to the original statistical spirit of the KTMAS, and, as such, it is a good candidate to approximate~\eqref{eqUrbanAsymmetricHomogeneousWeakFormBoltzmannType}.

\subsection{Emigration rules}
\label{secEmigrationRules}
We recall that, in our context, the agents are cities and interactions among them consist in exchanges of inhabitants due to emigration patterns. An emigration model can be specified by choosing particular forms for the functions $\psi,\,\psi_\ast$ in~\eqref{eqUrbanInteractionRules}, replacing the indices $i,\,i_\ast$ by $x,\,x_\ast$ consistently with the dense graph limit.

We consider the class of emigration models described by the rules
\begin{equation}
    s'=s\bigl(1-E(x,s,x_\ast,s_\ast)\bigr), \qquad s_\ast'=s_\ast+sE(x,s,x_\ast,s_\ast),
    \label{eqExplicitUrbanInteractionRules}
\end{equation}
partially inspired by~\cite{Gualandi2019SDC}. Here, $E$ is the \textit{emigration rate}.
            
\begin{rmk}
One could also define
$$ s'=s\bigl(1-E(x,s,x_\ast,s_\ast)+\eta\bigr), $$
where $\eta$ is a centred random variable, i.e. one with zero expectation, accounting for natural stochastic fluctuations in a city's population size due, for instance, to birth and death rates. Rules~\eqref{eqExplicitUrbanInteractionRules} just assume that the birth and death rates always compensate.
\end{rmk}
            
We further characterise $E$ by means of the following
\begin{hyp} \label{hypEmigrationRateLimits}
To ensure physically meaningful, i.e. non-negative, post-interaction states $s',\,s_\ast'$, we assume that there exists a constant $\bar E\in (0,\,1)$ such that
$$ 0\leq E(x,s,x_\ast,s_\ast)\leq\bar E, \qquad \forall s,\,s_\ast\in\R_+,\ \forall\,x,\,x_\ast\in [0,\,1]. $$
\end{hyp}

It is not difficult to see that rules~\eqref{eqExplicitUrbanInteractionRules} conserve the total population at each exchange of people between cities:
$$ s'+s_\ast'=s+s_\ast. $$
This is reasonable, since every emigration pattern from one city corresponds to an immigration pattern into another, leaving the combined population of the two interacting cities unchanged.

\section{Implementation and simulations of the kinetic model}
\label{secImplementationSimulation}
In this section, we consider several possible models for the emigration rate $E$ in~\eqref{eqExplicitUrbanInteractionRules} and we analyse the corresponding predictions of the Boltzmann-type equations on both the exact graph topology, cf.~\eqref{eqUrbanAsymmetricHomogeneousWeakFormBoltzmannType}, and the approximate random graph topology, cf.~\eqref{eqRelaxedGraphBoltzmannType}. Given the lack of well-established physical theories of emigration patterns, our main aim is to identify a minimal set of empirical factors that may influence the distribution of the population across cities, by comparing the model predictions with real-world data.

\subsection{Monte Carlo method}
\label{secMonteCarloMethod}
\begin{algorithm}[!t]
    \caption{Nanbu--Babovsky-type Monte Carlo algorithm for~\eqref{eqUrbanInteractionAlgorithm}}
    \label{algMonteCarlo}
    \KwData{%
        \(N\in\mathbb{N}_+\) (number of cities),
        \(\Delta t\leq 1\) (time step),
        \(T>0\) (final time),
        \(P\) (initial total population),
        \(\mathbf A=(a_{i,j})_{i,j=1}^{N}\) (adjacency matrix)%
    }
    \(\Xi\gets(\xi_1,\xi_2,\dots,\xi_N)=\mathbf{A}\mathbf{1}\)\;
    \(\mathcal{S}^0\gets(s_1^0,s_2^0,\dots,s_N^0)=\frac{P}{N}\mathbf 1\in\mathbb{R}_+^N\)\;\label{algMonteCarloDistribuzioneIniziale}
    \For{\(n=0,1,2,\dots,\left\lfloor T/\Delta t\right\rfloor-1\)}{
        \(p\gets\) random permutation of \(\{1,2,\dots,N\}\)\;
        \For{\(k=1,2,\dots,\left\lfloor N/2\right\rfloor\)}{
            \(i\gets p(k)\); \(i_\ast\gets p(\left\lfloor N/2\right\rfloor+k)\)\;
            \eIf{the exact graph topology is used}{
                sample \(\Theta\sim\operatorname{Bernoulli}(a_{i,i_\ast}\Delta t)\)\;
            }{
                sample \(\Theta\sim\operatorname{Bernoulli}\!\left(\frac{\xi_i\xi_{i_\ast}}{d_N}\Delta t\right)\)\;
            }
            \(x_i\gets\frac{\xi_i}{N}\)\;
            \(x_{i_\ast}\gets\frac{\xi_{i_\ast}}{N}\)\;
            \eIf{\(\Theta=1\)}{
                \(E\gets E(x_i,s_i^n,x_{i_\ast},s_{i_\ast}^n)\)\;
                \(s_i^{n+1}\gets s_i^n(1-E)\)\;
                \(s_{i_\ast}^{n+1}\gets s_{i_\ast}^n+s_i^nE\)\;
            }{
                \(s_i^{n+1}\gets s_i^n\)\;
                \(s_{i_\ast}^{n+1}\gets s_{i_\ast}^n\)\;
            }
        }
        \(\mathcal{S}^{n+1}\gets(s_1^{n+1},s_2^{n+1},\dots,s_N^{n+1})\)\;
    }
\end{algorithm}

To solve~\eqref{eqUrbanAsymmetricHomogeneousWeakFormBoltzmannType},~\eqref{eqRelaxedGraphBoltzmannType} numerically, we use a Nanbu--Babovsky-type Monte Carlo scheme~\cite{babovsky1986M2AS,nanbu1980JPSJ}, which consists in simulating directly the binary interactions described by the agent-based stochastic algorithm~\eqref{eqUrbanInteractionAlgorithm} supplemented by the interaction rules~\eqref{eqUrbanInteractionRules},~\eqref{eqExplicitUrbanInteractionRules}. We describe the scheme in detail in Algorithm~\ref{algMonteCarlo}.

We note that, in line~\ref{algMonteCarloDistribuzioneIniziale} of Algorithm~\ref{algMonteCarlo}, the array $\mathcal{S}^0$ of initial city sizes is prescribed deterministically by distributing a total population of size $P$ uniformly across the cities.
    
\subsection{Numerical results}
\label{secNumericalResults}
All real-world data used to inform the numerical simulations, as well as those used for comparison with the numerical results, were obtained from the Italian National Institute of Statistics (ISTAT). The data refer to $1991$ and to the Italian region of Sardinia, a choice primarily inspired by~\cite{DeMontis2007}. In more detail,~\cite{ISTATPendolarismo1991} provided the data to build the graph and its adjacency matrix, while~\cite{ISTATCensimento1991} provided the actual population distributions to which the numerical results were then compared.

\begin{table}[!t]
\centering
\caption{Data and parameters of Sardinia in $1991$~\cite{ISTATCensimento1991,ISTATPendolarismo1991}}
\begin{tabular}{cccccccc}
    \toprule
    $N$ & $P$ & $\beta$ & $\zeta$ & $\mu_1$ & $\sigma_1$ & $\mu_2$ & $\sigma_2$ \\
    \midrule
    $375$ & $1.648\cdot 10^6$ & $1.273$ & $0.929$ & $3.227$ & $0.408$ & $4.099$ & $0.556$ \\
    \bottomrule
    \end{tabular}
\label{tabSardiniaDataParameters}
\end{table}

The main data ($N$, $P$) of the Italian region of Sardinia are listed in Table~\ref{tabSardiniaDataParameters} along with the fitted parameters of the Pareto ($\beta$, cf.~\eqref{eqComplementaryCumulativePareto}) and bilognormal distributions ($\zeta$, $\mu_1$, $\sigma_1$, $\mu_2$, $\sigma_2$, cf.~\eqref{eqLognormalDistribution},~\eqref{eqBimodalLognormalDistribution}).
                        
Concerning the plots, their complex definitions are not necessary to understand the results shown hereafter. Hence, they are omitted here and deferred to Appendix~\ref{apxPlotDefinitions} for the interested reader. Nonetheless, three features common to all numerical tests have to be kept in mind:
\begin{enumerate}[label=(\roman*)]
\item Due to the stochastic nature of Algorithm~\ref{algMonteCarlo} and to the relative smallness of the number $N$ of cities considered, cf. Table~\ref{tabSardiniaDataParameters}, typically in each test $R=100$ simulations have been performed and subsequently averaged.
\item The time step is kept fixed to $\Delta t=10^{-2}$ while the number $N_t$ of time steps is variable depending on the desired final time $T$.
\item For the sake of brevity, only the best fitting configuration empirically found is analysed in each test.
\end{enumerate}

In the following, we consider various possible choices for the emigration rate $E$ to specify the interaction model~\eqref{eqExplicitUrbanInteractionRules}, and discuss the resulting outcomes.

\subsubsection{Size rule}
\label{secSER}
A first possibility consists in choosing
\begin{equation}
    E(s,s_\ast):=\lambda\frac{\left(s_\ast/s\right)^\alpha}{1+\left(s_\ast/s\right)^\alpha},
    \label{eqSizeEmigrationRule}
\end{equation}
independent of the latent variables $x,\,x_\ast$, where $\lambda\in (0,\,1)$, consistently with Assumption~\ref{hypEmigrationRateLimits}, and $\alpha\in\R_+$. This is a Hill function of order $\alpha$, partially inspired by~\cite{Gualandi2019SDC}, which is also largely applied in biomedical contexts.

The rationale behind~\eqref{eqSizeEmigrationRule} is that the larger the relative population size of the destination city compared to that of the origin city the larger the emigration rate.

\newsavebox{\stabularbox}
\begin{table}[!t]
\caption{Parameters of all the studied reference configurations}
\label{tabParmetersReferenceConfigurations}
\newcommand\subtabular[4]{
\sbox{\stabularbox}{\begin{tabular}{@{} #1 @{}}#2\end{tabular}}%
                    \begin{subtable}[b]{\wd\stabularbox}
                        \centering
                        \begin{tabular}{@{} #1 @{}}#2\end{tabular}
                        \caption{#3}
                        \label{#4}
                    \end{subtable}
}
\newcommand\tableindent{\hskip 0pt plus .4fill}
\tableindent
\subtabular{lcccc}{
\toprule
& $\lambda$ & $\alpha$ & $N_t$ & $R$ \\
\midrule
Stable & $0.1$ & $0.5$ & $10^7$ & $100$ \\
Unstable & $0.1$ & $1.5$ & $10^6$ & $10$ \\
\bottomrule
}
{Figures~\ref{figSERStableReferenceConfiguration},~\ref{figSERUnstableReferenceConfiguration}}{tabSERParameters}
\hfill
\subtabular{ccc}{
\toprule
$\lambda$ & $\alpha$ & $N_t$ \\
\midrule
$0.15$ & $0.63$ & $3\cdot 10^6$ \\
\bottomrule
}
{Figure~\ref{figSDERReferenceConfiguration}}{tabSDERParameters}
\hfill
\subtabular{ccc}{
\toprule
$\lambda$ & $\alpha$ & $N_t$ \\
\midrule
$0.18$ & $0.44$ & $3\cdot 10^6$ \\
\bottomrule
}
{Figure~\ref{figSSERReferenceConfiguration}}{tabSSERParameters}
\tableindent
\mbox{}
\end{table}
    
\begin{figure}[p]
  \centering
  \vspace*{-.5em}
  \tikzextfig{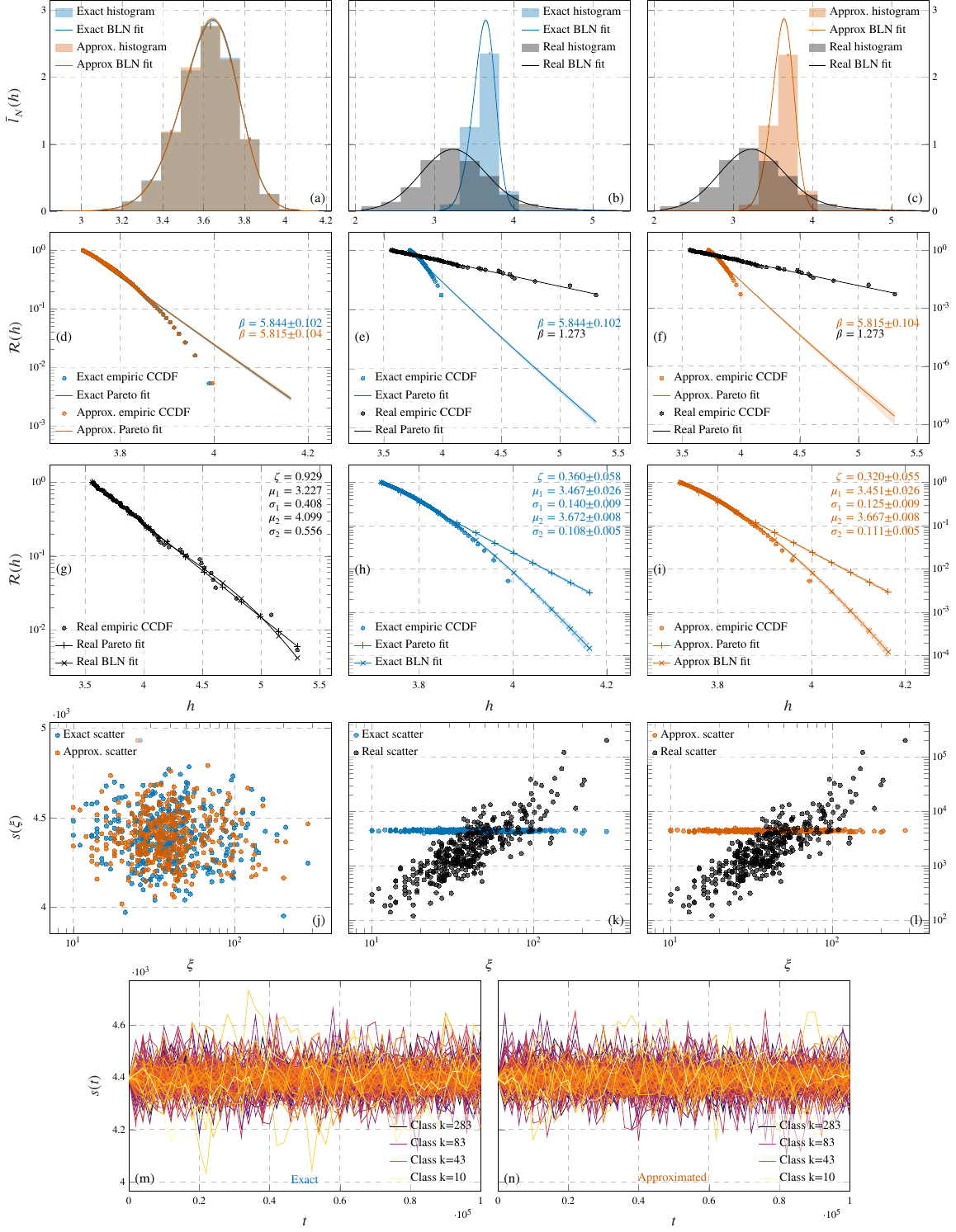}
  \newcaption{Study of the \textit{stable} reference configuration of~\eqref{eqSizeEmigrationRule} with parameters from Table~\ref{tabSERParameters}}{.6\linewidth}
  \label{figSERStableReferenceConfiguration}
\end{figure}

\begin{figure}[p]
  \centering
  \vspace*{-1em}
  \tikzextfig{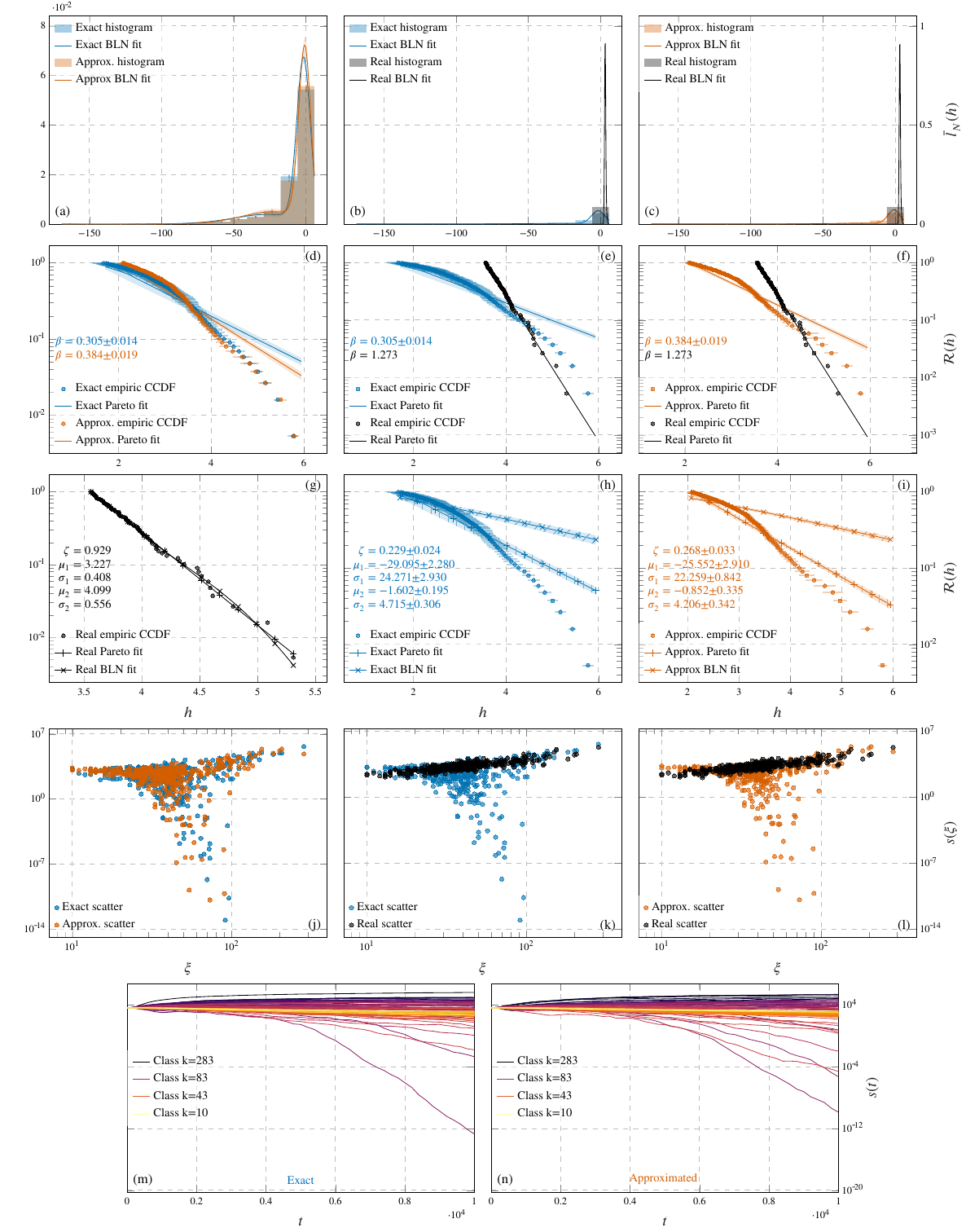}
  \newcaption{Study of the \textit{unstable} reference configuration of~\eqref{eqSizeEmigrationRule} with parameters from Table~\ref{tabSERParameters}}{.62\linewidth}
  \label{figSERUnstableReferenceConfiguration}
\end{figure}

We studied two reference configurations described by the parameters in Table~\ref{tabSERParameters}. The corresponding results are illustrated in Figures~\ref{figSERStableReferenceConfiguration},~\ref{figSERUnstableReferenceConfiguration}. These describe two trends: a stable one, which is not distant from the initial uniform distribution, and an unstable one, in which, on average, the medium-size cities depopulate.

We stress that the latter conclusion holds on average, since in single realisations it is much more likely that the larger cities depopulate, whereas over multiple realisations at least one with a large size manifests, thereby balancing the average. This is also the reason why only $10$ runs were considered instead of the usual $100$ in Table~\ref{tabSERParameters}, given that the average tends to mask the depopulation, which is nonetheless inevitable albeit slower. In addition to Figures~\ref{figSERUnstableReferenceConfiguration}j,l, the histograms in Figures~\ref{figSERUnstableReferenceConfiguration}a,c also confirm this tendency. Specifically, their tails elongate towards negative values, corresponding to the exponents of the cities that depopulate.
    
More generally, we have verified, albeit only numerically, that for every $\lambda>0$ there exists $\bar\alpha\in\R_+$ such that $\alpha<\bar\alpha$ results in a stable configuration while $\alpha\geq\bar\alpha$ yields an unstable one.

\subsubsection{Size-degree rule}
\label{secSDER}
To improve~\eqref{eqSizeEmigrationRule}, we introduce the relative degree ratio:
\begin{equation}
    E(x,s,x_\ast,s_*):=\lambda\frac{\left[\left(s_\ast/s\right)\cdot \left(x_\ast/x\right)\right]^\alpha}{1+\left[\left(s_\ast/s\right)\cdot \left(x_\ast/x\right)\right]^\alpha}.
    \label{eqSizeDegreeEmigrationRule}
\end{equation}
The rationale is that the more populous and \textit{better connected} the destination city $(x_\ast,\,s_\ast)$ is, the more attractive it is relative to the origin city $(x,\,s)$.

\begin{figure}[p]
  \centering
  \tikzextfig{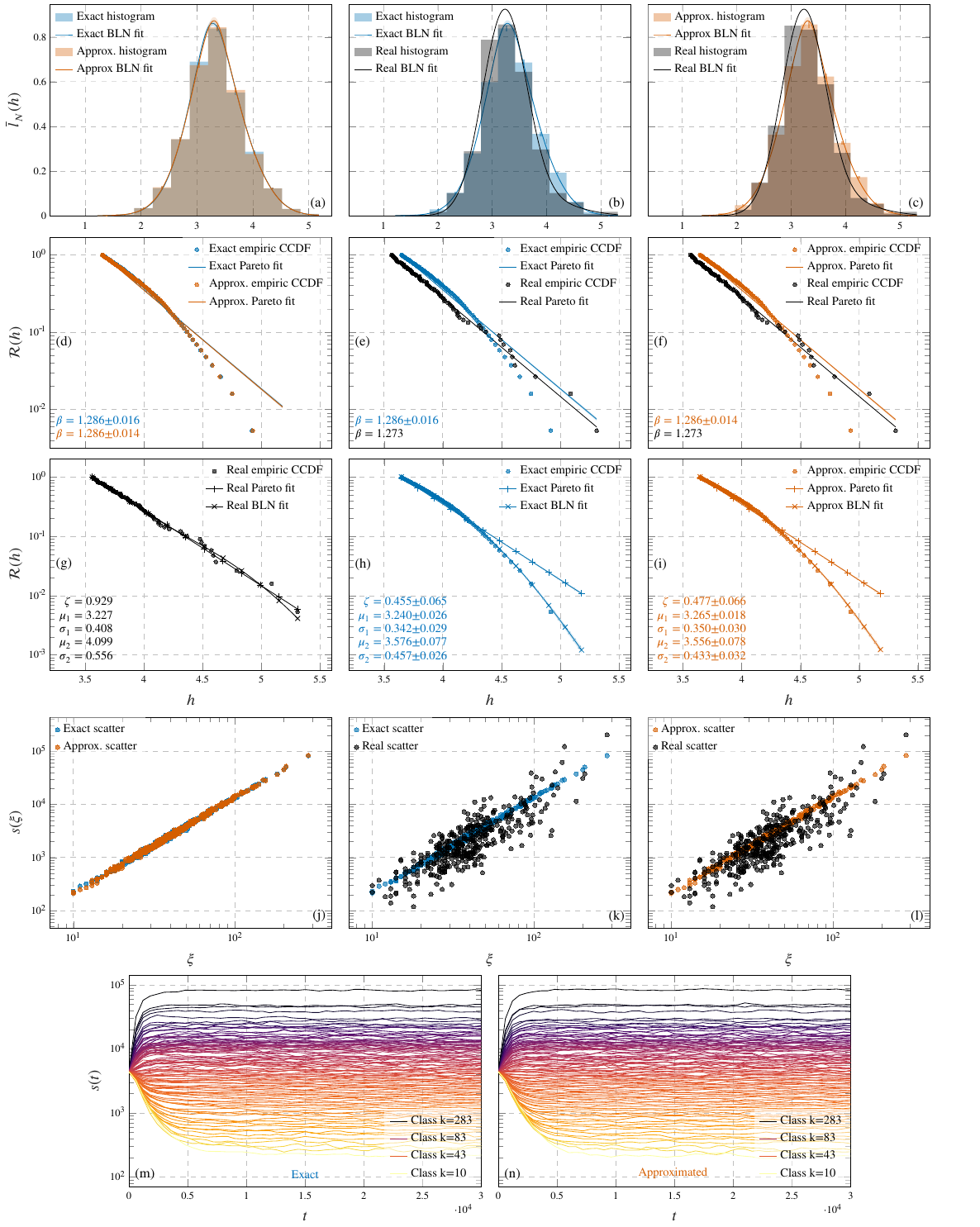}
  \newcaption{Study of the reference configuration of~\eqref{eqSizeDegreeEmigrationRule} with parameters from Table~\ref{tabSDERParameters}}{.58\linewidth}
  \label{figSDERReferenceConfiguration}
\end{figure}

The studied reference configuration is described by the parameters in Table~\ref{tabSDERParameters}, while the results are illustrated in Figure~\ref{figSDERReferenceConfiguration}. Overall, these results improve those obtained using~\eqref{eqSizeEmigrationRule} in several aspects:
\begin{itemize}
\item Figures~\ref{figSDERReferenceConfiguration}a, d, j show very good agreement between the results obtained from the exact graph topology and its random-graph-approximation.
\item In Figures~\ref{figSDERReferenceConfiguration}b, c, the simulated distributions are fairly similar to the real ones; in Figures~\ref{figSDERReferenceConfiguration}e, f the Pareto indices are reasonably close.
\item In Figures~\ref{figSDERReferenceConfiguration}g--i, the bilognormal profile follows the terminal part of the tail better than the Pareto profile.
\end{itemize}

On the other hand, in Figures~\ref{figSDERReferenceConfiguration}k, l we notice that the scatter plots present a similar positive correlation, albeit much tighter than the real one. It is nearly a one-to-one correlation, a clearly quite unrealistic outcome.

\subsubsection{Size-strength rule}
\label{secSSER}
In~\eqref{eqSizeDegreeEmigrationRule}, we may interpret the latent variable $x$ as the \textit{strength} of a city, i.e. the total number of commuters associated with that city, instead of the number of connections. This way, the same emigration rule~\eqref{eqSizeDegreeEmigrationRule} can be used to model the fact that the more populous and \textit{heavily trafficked} the destination city is, the more attractive it is relative to the origin city.

\begin{figure}[p]
  \centering
  \tikzextfig{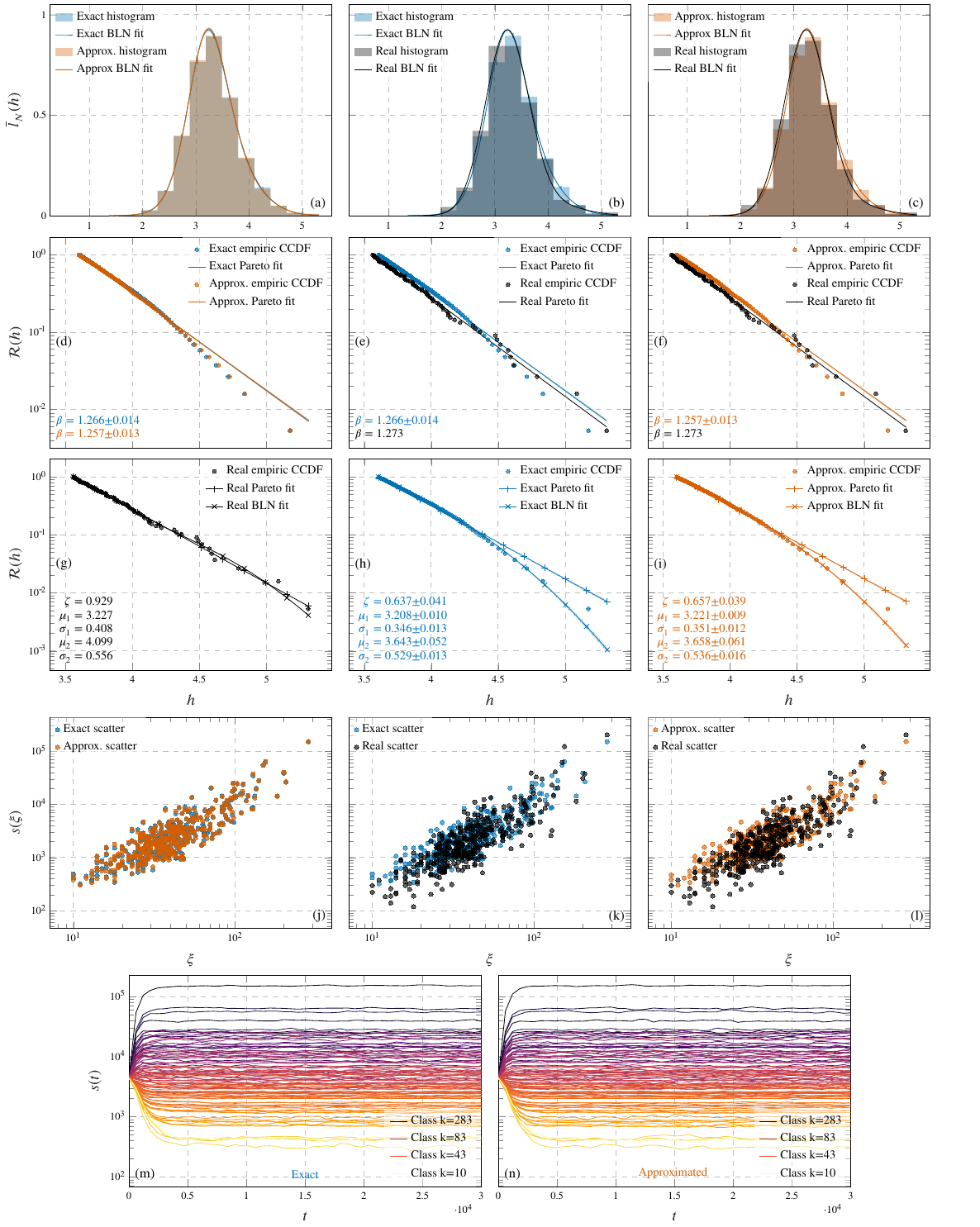}
  \newcaption{Study of the reference configuration of~\eqref{eqSizeDegreeEmigrationRule}, in which the latent variable $x$ is reinterpreted as the strength of a city, with parameters from Table~\ref{tabSSERParameters}}{.577\linewidth}
  \label{figSSERReferenceConfiguration}
\end{figure}

The reference configuration analysed here is now described by the parameters listed in Table~\ref{tabSSERParameters}, while the new results are shown in Figure~\ref{figSSERReferenceConfiguration}. Note that, in this case, we are not changing the mathematical model itself, but rather the type of real-world data used to validate its results. Indeed, these results are the best according to several criteria:
\begin{itemize}
\item Figures~\ref{figSSERReferenceConfiguration}a, d, j confirm that the simulation outcomes obtained using the random-graph approximation of the graph topology correspond almost exactly to those obtained using the exact graph topology.
\item In Figures~\ref{figSSERReferenceConfiguration}b, c, the simulated distributions are almost identical to the real one, and the relative Pareto indices in Figures~\ref{figSSERReferenceConfiguration}e, f also correspond.
\item Figures~\ref{figSSERReferenceConfiguration}h, i do not deviate significantly from the previous configuration.
\item The scatter plots in Figures~\ref{figSSERReferenceConfiguration}k, l show a larger spread than in the previous case while maintaining the same positive correlation with the real-world data. This captures correctly the size differentiation for a given degree.
\end{itemize}
    
The only drawback is visible in Figures~\ref{figSSERReferenceConfiguration}k, l: the emigration rule fails to correctly predict, on average, the city sizes in the lower portion of the real-world scatter plot. Specifically, the model prediction overestimates the size of the smaller cities and slightly underestimates that of the larger cities. Yet, on second thought, this is an issue already present in the previous case, cf. the slope of the line described by the scatter plot in Figures~\ref{figSDERReferenceConfiguration}k, l. We believe that it is possibly caused by the absence of intra-urban commuters in the reinterpretation of $x$ as the strength of a city, which considers only inter-urban commuters. Including also this aspect may significantly increase the strength of the medium-sized and larger cities and only marginally that of the smaller ones, thereby presumably correcting the slope.

\subsection{Interpretations of the results}
\label{secInterpretations}
In this section, we discuss the previous results in more detail, with the aim of identifying the microscopic, i.e. city-level, factors that are most likely to influence migration.
        
\subsubsection{Size rule}
From the results of~\eqref{eqSizeEmigrationRule}, it is clear that the relative population size alone is insufficient to accurately describe the city-size distribution. Indeed, if $s\leq s_\ast$, the destination city will invariably attract people from the origin city (and vice versa). Consequently, as soon as a city loses even a small portion of its population relative to all its neighbouring cities, it is bound to continue depopulating. This mechanism can explain the depopulation, \textit{on average}, of medium-sized cities: large cities draw population away from medium-sized cities, making the latter relatively unattractive vertices from the outset. In the short term, medium-sized cities depopulate to such an extent that they become relatively unattractive even compared to small cities. The latter do not interact directly with large cities and therefore do not initially experience the same depopulation as medium-sized cities. The outcome of these dynamics is shown in Figures~\ref{figSERUnstableReferenceConfiguration}k, l, at least in the unstable case.

\begin{figure}[t]
  \centering
  \tikzextfig{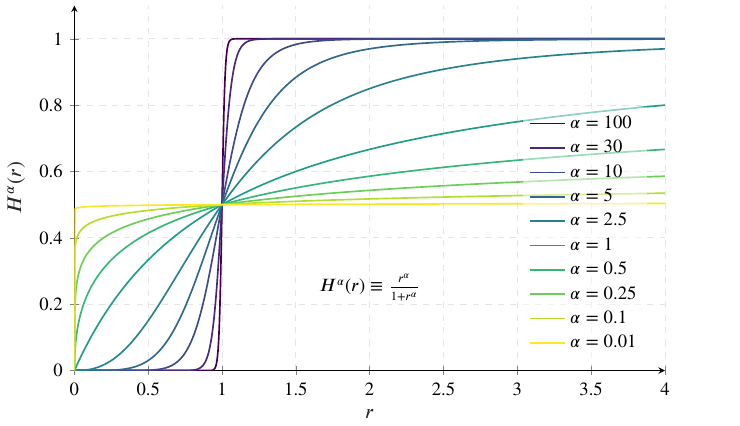}
  \newcaption{Hill function for different values of the degree $\alpha$.}{.775\linewidth}
  \label{figDegreeAHillFunction}
\end{figure}

Conversely, in the stable case, the interpretation is less straightforward. Certainly, $\alpha$ is not sufficiently small for the Hill function in~\eqref{eqSizeEmigrationRule} to become approximately constant (cf. Figure~\ref{figDegreeAHillFunction}) and equal to $\lambda/2$, particularly given that~\eqref{eqSizeDegreeEmigrationRule}, for the reference configuration in Table~\ref{tabSSERParameters}, yields an even lower value of $\alpha$. We can therefore only conclude that the issue lies in the emigration rule itself.

Overall, these results suggest that migration most likely does not depend \textit{solely} on the relative population sizes of the interacting cities.
            
\subsubsection{Size-degree rule}
The results obtained with the emigration rule~\eqref{eqSizeDegreeEmigrationRule} are considerably more realistic. Nevertheless, unlike relative population size, the introduction of relative degree through the ratio $x_\ast/x$ is not immediately intuitive. In other words, one might wonder why it is reasonable to assume that a city with a low degree, i.e. one that is less connected, should also have a small population. There is a risk that the model is being forced towards results that, although consistent with the real-world data, are not realistic in terms of the underlying mechanisms. This modelling choice can, however, be motivated by two naturally observable constraints.
\begin{itemize}
\item The first is a physical constraint arising from the spatial nature of the graph. By their nature, edges occupy the spaces between cities and connect to them along portions of their boundary. Therefore, a city can increase its degree mainly by expanding so as to have a larger boundary, which in turn requires a larger population to sustain it.
\item The second is a social constraint arising from commuter flows: the greater the connectivity, the more complex the transport network, and such complexity can be justified only if there are sufficiently many commuters using it. Such commuter flows, in turn, are more likely to be attracted to a city with a sufficiently large population to sustain them.
\end{itemize}

\begin{figure}[t]
  \centering
  \tikzextfig{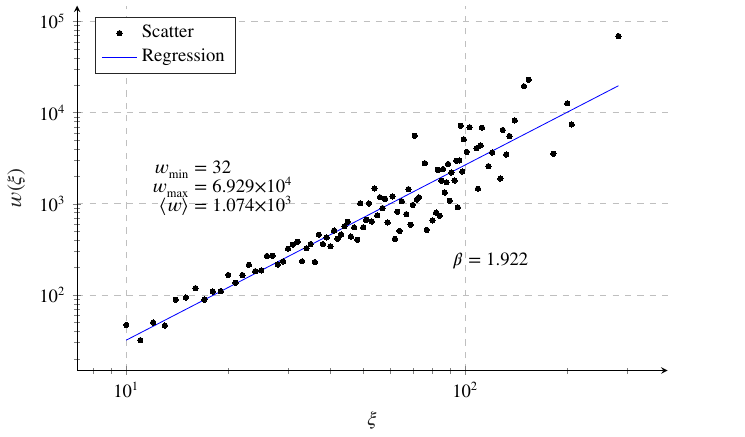}
  \newcaption{Strength vs degree of the Sardinian network}{.75\linewidth}
  \label{figStrengthVsDegree20}
\end{figure}

The social constraint is indeed supported by Figure~\ref{figStrengthVsDegree20}, which shows a positive correlation between degree and strength.
                
The physical constraint justifies this type of interaction rule because the graph is inherently spatial. Similarly, the social constraint supports this interaction rule provided that degree and strength are positively correlated. Together, these considerations suggest the relationship $s\propto x$, which is also supported by the empirical scatter shown in Figures~\ref{figSDERReferenceConfiguration}k, l. This relationship makes the introduction of the ratio $x_\ast/x$ plausible from a modelling perspective.

It is also worth considering why the emigration rule~\eqref{eqSizeDegreeEmigrationRule} leads the system to a non-trivial equilibrium configuration at long times. In particular, one may wonder why the population flow ``small cities $\to$ medium-sized cities $\to$ large cities'' does not occur, as this would ultimately lead to the depopulation of small cities. This can only be explained by the presence of a backflow naturally induced by~\eqref{eqSizeDegreeEmigrationRule}, which balances the flow towards relatively more populous and better-connected cities. In other words, introducing degree in~\eqref{eqSizeEmigrationRule} not only makes medium-sized cities more attractive than small cities in the short term, but also stabilises the long-term, i.e. stationary, interactions. However, the precise mechanism producing this stabilising reverse flow remains unclear, since the ratio $r$ in the Hill function (cf. Figure~\ref{figDegreeAHillFunction}) should be smaller in~\eqref{eqSizeDegreeEmigrationRule} than in~\eqref{eqSizeEmigrationRule}.
                
In conclusion, these results suggest that, according to~\eqref{eqSizeDegreeEmigrationRule}, migration is likely to be governed by \textit{at least} two factors: the relative population sizes and the connectivity of the interacting cities.
            
\subsubsection{Size-strength rule}
Reinterpreting the latent variables $x$ and $x_\ast$ in~\eqref{eqSizeDegreeEmigrationRule} as the strengths of the origin and destination cities makes it possible to distinguish between cities with the same degree but not necessarily the same commuter traffic. This natural distinction allows us to relax the almost one-to-one relationship between degree and population size observed in Figures~\ref{figSDERReferenceConfiguration}k, l, replacing it with the more realistic positive correlation shown in Figures~\ref{figSSERReferenceConfiguration}k, l.
            
Since these results provide the best agreement both qualitatively and quantitatively, we conclude that the migration is \textit{most likely} governed by the relative population sizes and strengths of the interacting cities.
            
\subsubsection{The roles of~\texorpdfstring{$\boldsymbol{\lambda}$}{} and~\texorpdfstring{$\boldsymbol{\alpha}$}{}}
\begin{table}[!t]
\centering
\caption{Parametric studies in the Supplementary Material}
\label{tabRETFParametricStudies}
\begin{tabular}{lcccc}
\toprule
& $N_p$ & $\lambda$ & $\alpha$ & $N_t$ \\
\midrule
Parametric study on $\lambda$: & $5$ & $[0.01,\,0.3]$ & $0.44$ & $10^7$ \\
Parametric study on $\alpha$: & $5$ & $0.18$ & $[0.3,\,0.6]$ & $3\cdot 10^6$ \\
\bottomrule
\end{tabular}
\end{table}
                
All the proposed emigration rules involve the parameters $\lambda$ and $\alpha$. To investigate their roles, we performed two parametric studies, which are presented in the Supplementary Material for conciseness. Both studies are based on the reference configuration for the size-strength rule~\eqref{eqSizeDegreeEmigrationRule} reported in Table~\ref{tabSSERParameters}, with only one of the two parameters varied at a time. Table~\ref{tabRETFParametricStudies} summarises the main settings adopted for these parametric studies, where $N_p=5$ denotes the number of values uniformly sampled from the specified intervals for each parameter.

\begin{figure}[p]
  \centering
  \tikzextfig{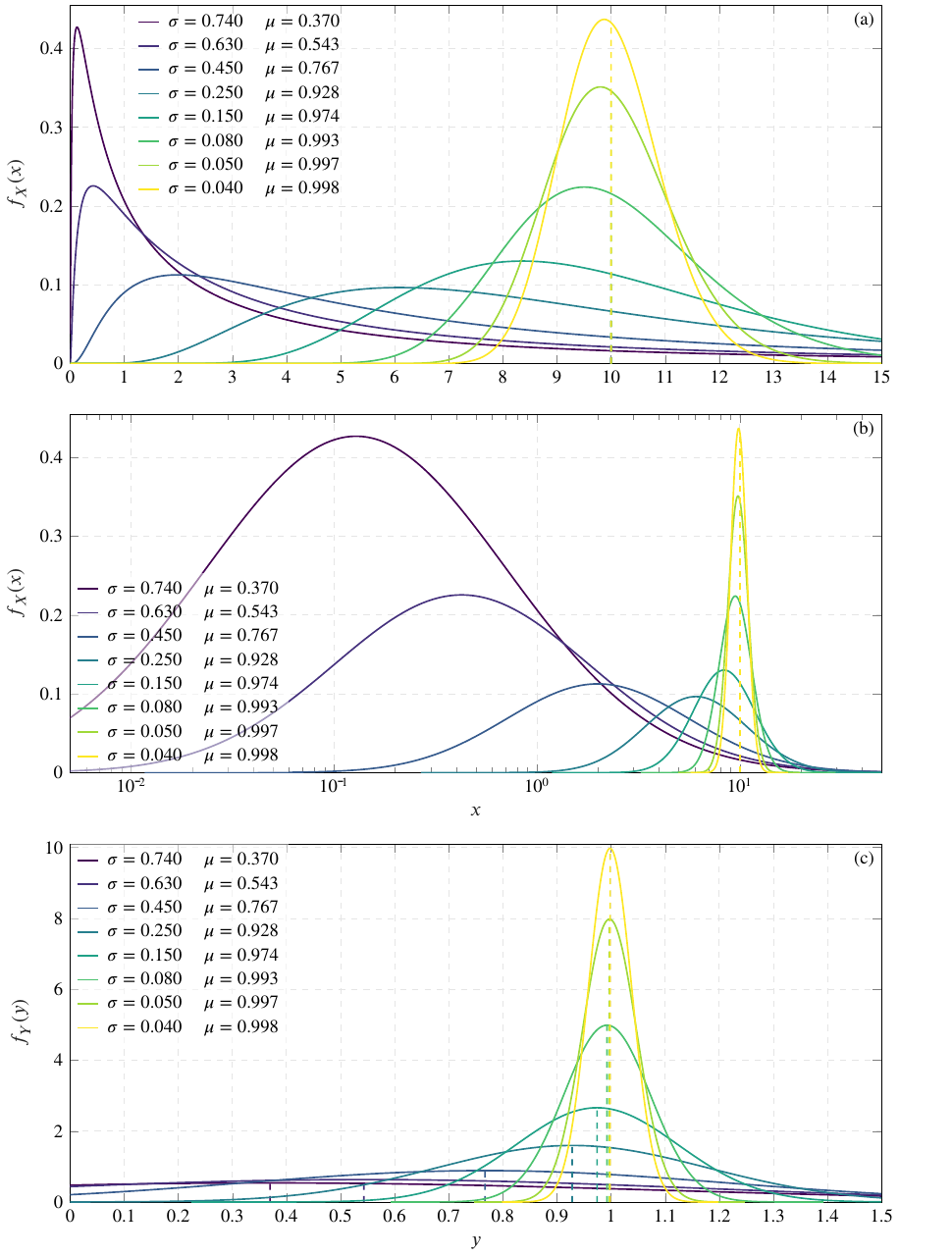}
  \newcaption{Probability density function of a random variable $X\sim\operatorname{Lognormal}(\mu,\,\sigma)$ for different parameters values, with the expected value held constant at $\mathbb{E}[X]=10^{\mu+\sigma^2/2}=10$. The distributions are shown on (a) linear, (b) logarithmic, and (c) log--log scales. Dashed lines indicate the expected value of each distribution}{.65\linewidth}
  \label{figLognormalFunctionExample}
\end{figure}

One effect of $\lambda$ and $\alpha$ is that they influence the parameters $\mu$ and $\sigma^2$ of the lognormal distribution~\eqref{eqLognormalDistribution}. This is because, owing to the conservation of the total population in each binary exchange between cities, the sample mean, fixed initially as $P/N$ in Algorithm~\ref{algMonteCarlo}, cf. line~\ref{algMonteCarloDistribuzioneIniziale}, remains constant over time. Consequently, the mean of the lognormal distribution must also remain constant, i.e. $10^{\mu+\sigma^2/2}=P/N$, independently of the values of $\lambda$ and $\alpha$. This, in turn, implies that $\mu$ and $\sigma$ vary with $\lambda$ and $\alpha$. This behaviour is difficult to discern on either a linear or a logarithmic scale, but becomes more apparent on a log--log scale, cf. Figure~\ref{figLognormalFunctionExample}, since $\mu$ and $\sigma$ are precisely the mean and standard deviation of the corresponding Gaussian distribution after logarithmic transformation. Similar reasoning applies to the bilognormal fit~\eqref{eqBimodalLognormalDistribution}, since it is a convex combination of two lognormal distributions.

In more detail, the parametric study in which $\lambda$ is varied while $\alpha$ is held fixed leads to the following conclusions:
\begin{itemize}
\item As $\lambda$ increases, the variance increases, thereby broadening the distribution, albeit only slightly.
\item The Pareto index remains \textit{de facto} almost constant, with an average variation of only approximately $0.05$.
\item The relationship between the Pareto and bilognormal distributions remains unchanged: the latter provides a better approximation of the far end of the tail, whereas the two distributions coincide near its beginning.
\item The degree scatter remains essentially unchanged as $\lambda$ varies, and the average behaviour is the same.
\item All considered cases converge to a steady state, with the convergence time decreasing as $\lambda$ increases.
\end{itemize}
We further observe that all the variations described above are small, despite $\lambda$ increasing by as much as an order of magnitude. We therefore conclude that $\lambda$ has little influence on the results, except for the convergence time. Nonetheless, for $\lambda>0.3$, the system reaches an unstable configuration.

Conversely, the parametric study in which $\alpha$ is varied while $\lambda$ is held fixed leads to the following conclusions:
\begin{itemize}
\item As $\alpha$ increases, the variance increases significantly, thereby reducing the mean and broadening the distribution.
\item The Pareto index varies significantly, decreasing by nearly an order of magnitude.
\item The relationship between the Pareto and bilognormal distributions in terms of their approximation of the tail remains unchanged.
\item For small values of $\alpha$, the degree scatter leads to an overestimation of the size of the minor and medium-sized cities and an underestimation of those of large cities; as $\alpha$ increases, this tendency is reversed.
\item All considered cases converge to a steady state, with the convergence time increasing as $\alpha$ increases.
\end{itemize}
In this case, all the variations described above are significant. In particular, the Pareto index appears to be strongly affected by the value of $\alpha$. Similarly, for $\alpha>0.6$, the system reaches an unstable configuration.

Overall, in light of these parametric studies, we propose the following interpretations of the roles of $\lambda$ and $\alpha$:
\begin{itemize}
\item $\lambda$ is an \textit{attraction} parameter: the greater its value, the larger the fraction of the population that large cities can attract from smaller cities in a single interaction.
\item $\alpha$ is a \textit{repulsion} parameter: the greater its value, the less likely people from small cities are to move towards large cities. This effect is clearly inversely related to the magnitude of the stabilising backflow.
\end{itemize}
Alternatively, these roles can be interpreted in terms of the Hill function (cf. Figure~\ref{figDegreeAHillFunction}): $\lambda$ mainly influences the part of the function corresponding to $r>1$, and hence the attraction, whereas $\alpha$ mainly influences the part corresponding to $r\leq 1$, and hence the repulsion.

\section{Conclusions}
\label{secConclusions}
The results presented in this paper support the vertex-agent perspective as a modelling framework in an urban context, both for the city-size distribution and for its Pareto index. The random-graph approximation introduced in Definition~\ref{defDegreeMatrix} produces results that closely match those obtained using the exact graph topology. All the interaction rules faithfully reproduce both the empirical distribution and the Pareto index of its tail. However, the bimodal lognormal distribution provides a better fit to the tail than the Pareto distribution, confirming the findings of~\cite{Gualandi2019SDC,Gualandi2019HBLD}. Both fits have advantages and disadvantages:
\begin{itemize}
\item From an applied standpoint, the Pareto distribution is more straightforward to fit. Its single parameter is essentially uniquely determined by the data, ensuring that the fitting procedure consistently yields the same result.
\item From a mathematical standpoint, the bimodal lognormal distribution is more appealing. Nevertheless, its greater flexibility comes at the cost of five parameters, whose estimation may yield different results across fitting attempts. Moreover, two of these parameters are particularly difficult to compare across different fits.
\end{itemize}

The work presented here is by no means exhaustive and opens up several avenues for future research. In particular, possible developments include:
\begin{itemize}
\item Incorporating intra-urban commuting into the size-strength rule~\eqref{eqSizeDegreeEmigrationRule}, or considering interaction rules involving other graph parameters, such as betweenness centrality or the clustering coefficient.
\item Deriving the Fokker--Planck approximation of~\eqref{eqUrbanAsymmetricHomogeneousWeakFormBoltzmannType} as a basis for the analytical study of the stationary distribution.
\item Relaxing the simplifying Assumption~\ref{hypStaticGraph} of a non-coevolving network in order to model dynamic connections that evolve alongside population migration, with a possible comparison with empirical data, where available.
\item Investigating whether the results can be reproduced for European or international networks, thereby assessing the validity of the present theory at larger scales.
\end{itemize}

\appendix
    
\section{Results reproducibility}

\subsection{Technical specifications}
The code used to perform the numerical simulations is available in the repository~\cite{Taralli2026}. It was written entirely in Python 3.13.7, and the simulations were run on a Windows-based machine equipped with an Intel Core i5-7500 processor (quad-core, 3.40 GHz).

\subsection{Confidence interval}
Let $R\in\N_+$ denote the number of simulations performed, and let $\mathbf S\in\R_+^R$ be the random vector whose components represent the population size of a city. By construction of Algorithm~\ref{algMonteCarlo}, the components $S_1,\,S_2,\,\dots,\,S_R$ of $\mathbf S$ are independent and identically distributed, which allows us to define
$$ \bar S:=\frac{1}{R}\sum_{r=1}^{R}S_r, \qquad V:=\frac{1}{R-1}\sum_{r=1}^R{(S_r-\bar S)}^2 $$
as the sample mean and variance, respectively, of the $S_r$'s, $r=1,\,\dots,\,R$. If $\mu\in\R$ is the true mean of the distribution of the $S_r$'s, Student's \(T\)-distribution with parameter $R-1$ can be written as
$$ T:=\frac{\bar S-\mu}{\sqrt{\frac{V}{R}}}. $$
From this, we can define the (symmetric) confidence interval with significance level $\delta\in (0,\,1)$ by setting
$$ \operatorname{Prob}\!\left(-t_{R-1}^{\delta/2}\leq T\leq t_{R-1}^{\delta/2}\right)=1-\delta, $$
where $t_{R-1}^{\delta/2}\in\R$ is such that
$$ \operatorname{Prob}\!\left(T<t_{R-1}^{\delta/2}\right)=1-\frac{\delta}{2}. $$

Expanding $T$ and isolating the true mean $\mu$, we obtain
$$ \operatorname{Prob}\!\left(\bar{S}-t_{R-1}^{\delta/2}\sqrt{\frac{V}{R}}\leq\mu\leq\bar{S}+t_{R-1}^{\delta/2}\sqrt{\frac{V}{R}}\right)=1-\delta, $$
whence
\begin{equation}
    \IC^\delta_R(\mathbf S):=\left[\bar{S}-t_{R-1}^{\delta/2}\sqrt{\frac{V}{R}},\,\bar{S}+t_{R-1}^{\delta/2}\sqrt{\frac{V}{R}}\right]
    \label{eqConfidenceInterval}
\end{equation}
is the interval estimator defining the desired confidence interval.

The confidence interval essentially represents the statistical error one is prepared to accept: it assesses how likely the interval estimator~\eqref{eqConfidenceInterval} is to contain the parameter $\mu$. In other words, $\IC^\delta_R$ quantifies the uncertainty associated with estimating the mean: for a given sample $\mathbf s$ of the random vector $\mathbf S$, the wider the confidence interval, the greater the uncertainty with which the sample mean $\bar s$ estimates the true mean $\mu$. Conversely, the narrower the confidence interval, the more precise the estimate $\bar s$, in the sense that $\mu$ lies within a small neighbourhood of the sample mean. From this perspective, the confidence interval is conceptually analogous to the precision of a measuring instrument.

In all the tests presented in Section~\ref{secImplementationSimulation}, we set $\delta=0.05$, corresponding to a confidence level of $95\%$.

The interval estimate can then be obtained directly from the interval estimator~\eqref{eqConfidenceInterval} by considering the realisations $\mathbf s$ of the random vector $\mathbf S$, which are precisely the outcomes of the $R$ numerical simulations.
        
\subsection{Data structure}
The data are organised into a fourth-order tensor $\mathbf s=(s^{n,a}_{i,r})\in\R_+^{N_f\times 2\times N\times R}$, whose indices have the following meanings:
\begin{itemize}
\item $n$ denotes the time index;
\item $a$ denotes the type of simulation ($a=1$ for the exact graph topology; $a=2$ for the approximated graph topology);
\item $i$ denotes the city index;
\item $r$ denotes the simulation number.
\end{itemize}

The time index is further characterised by three parameters in $\N_+$:
\begin{itemize}
\item $N_t:=\left\lfloor\frac{T}{\Delta t}\right\rfloor$ is the total number of simulated time steps;
\item $N_s\ll N_t$ is the number of snapshots retained from the simulated time steps;
\item $N_f<N_s$ is the number of reduced time points obtained from the snapshots.
\end{itemize}

Both the snapshots and the reduced time points are sampled at evenly spaced intervals\footnote{A more refined discretisation would use an adaptive time step based on the plots discussed in Section~\ref{secEvolutionMeanCitySizeByDegreeClass}.}, with steps $\Delta s:=N_t/N_s$ and $\Delta f:=N_s/N_f$, respectively. For simplicity, we assume that $N_t$ is a multiple of $N_s$ and that $N_s$ is a multiple of $N_f$. We make an analogous assumption also for $T$ and $\Delta t$.

Using Algorithm~\ref{algMonteCarlo}, a total of $N_t$ times with time steps of size $\Delta t$ are simulated, of which $N_s$ are retained and stored in the snapshot tensor $\undertilde{\mathbf s}=(\undertilde s^{n,a}_{i,r})\in\R_+^{N_s\times 2\times N\times R}$. The tensor $\undertilde{\mathbf s}$ is then averaged over the time index according to
$$ s^{n,a}_{i,r}:=\frac{1}{\Delta f}\sum_{j=1}^{\Delta f}\undertilde s^{(n-1)\Delta f+j,a}_{i,r}, $$
so that each consecutive block of $\Delta f$ snapshots is replaced by its average. This smoothing reduces the noise in $\mathbf s$ relative to $\undertilde{\mathbf s}$, thereby improving the readability of the resulting plots.

Three notions of physical time can be defined, depending on the interpretation of the index $n$:
\begin{itemize}
\item $t_n:=n\Delta{t}$, $n=1,\,\dots,\,N_t$, is the discretised time;
\item $t^s_n:=t_n\Delta{s}$, $n=1,\,\dots,\,N_s$, is the snapshot time;
\item $t^f_n:=t^s_n\Delta{f}$, $n=1,\,\dots,\,N_f$, is the reduced time.
\end{itemize}
In each case, $t_{N_t}=t^s_{N_s}=t^f_{N_f}=T$.

In all the tests presented in Section~\ref{secImplementationSimulation}, we set $\Delta t=10^{-2}$, $N_s=1000$, and $N_f=50$. Moreover, we choose $N_t$ such that the final simulated time $T$ is slightly greater than the convergence time.

In most plots, only the distribution at the final time $T$ is considered, \textit{without} time averaging. For ease of notation, the resulting third-order tensor $\mathbf s^T:=\undertilde{\mathbf s}^{N_s}$ is therefore denoted, with a slight abuse of notation, by $\mathbf s^T\in\R_+^{2\times N\times R}$.
        
\subsection{Plot definitions}
\label{apxPlotDefinitions}
\subsubsection{Histograms}
The first rows of Figures~\ref{figSERStableReferenceConfiguration}--\ref{figSSERReferenceConfiguration} show the normalised histograms of the city-size distribution at the final time $T$.

We first consider the minimum and maximum entries of the tensor $\mathbf s^T$:
$$ s_{\min}:=\min_{a,\,i,\,r}s^{T,a}_{i,r}, \qquad s_{\max}:=\max_{a,\,i,\,r}s^{T,a}_{i,r}, $$
and use them to construct an evenly spaced grid defining the histogram bins. Let $N_c$ be the number of such bins. We then introduce the function $\mathcal{B}:\R^N\to\R^{N_c}$, which returns the normalised bin values of the histogram, and define the tensor $\mathbf b\in\R_+^{2\times N_c\times R}$ by $\mathbf b^a_{\cdot,r}:=\mathcal{B}(\mathbf s^{T,a}_{\cdot,r})$ for every $a$ and $r$. Finally, for every $a$ and $c$, we evaluate the confidence intervals
$$ \IC_R^\delta(\mathbf b^a_c)=\left[\bar{b}^a_c-t^{\delta/2}_{R-1}\sqrt{\frac{v^a_c}{R}},\,\bar{b}^a_c+t^{\delta/2}_{R-1}\sqrt{\frac{v^a_c}{R}}\right]. $$
We then plot the mean histograms for both the ``exact'' ($a=1$) and ``approximated'' ($a=2$) simulations using the mean vectors $\bar{\mathbf b}^a$, together with the associated statistical uncertainty in each bin, represented by adding and subtracting the vector $t^{\delta/2}_{R-1}\sqrt{\frac{\mathbf v^a}{R}}$.

\subsubsection{Bimodal lognormal distribution}
The first rows of Figures~\ref{figSERStableReferenceConfiguration}--\ref{figSSERReferenceConfiguration} also show the bimodal lognormal distributions fitted to the data represented by the histograms at the final time $T$.

To represent these distributions, we follow a procedure similar to that used for the histograms. Let $\mathcal{L}(\cdot;\,\mathbf s^{T,a}_{\cdot,r}):\R_+\to\R_+$ denote the probability density of a bilognormal random variable, cf.~\eqref{eqBimodalLognormalDistribution}, fitted to the vector $\mathbf s^{T,a}_{\cdot,r}$. Moreover, let $\boldsymbol{\mathcal{L}}(\cdot;\,\mathbf s^{T,a}):\R_+\to\R_+^R$ be the vector-valued function defined by
$$ \boldsymbol{\mathcal{L}}(y;\,\mathbf s^{T,a}):=
    {\left[\mathcal{L}(y;\mathbf s^{T,a}_{\cdot,1}),\,\mathcal{L}(y;\,\mathbf s^{T,a}_{\cdot,2}),\,\dots,\,\mathcal{L}(y;\,\mathbf s^{T,a}_{\cdot,R})\right]}^\top, $$
which collects the fits obtained from all $R$ simulations into a single vector. The corresponding confidence intervals are then given by
$$ \IC_R^{\delta}(\boldsymbol{\mathcal{L}}(y;\,\mathbf s^{T,a}))=
    \left[\bar{\mathcal L}^a(y)-t^{\delta/2}_{R-1}\sqrt{\frac{v^a(y)}{R}},\,\bar{\mathcal L}^a(y)+t^{\delta/2}_{R-1}\sqrt{\frac{v^a(y)}{R}}\right],
        \qquad y\in [s_{\min},\,s_{\max}], $$
where, as before, $\bar{\mathcal L}^a(y)$ denotes the mean function, while $\pm t^{\delta/2}_{R-1}\sqrt{\frac{v^a(y)}{R}}$ represents its statistical uncertainty.

Since $y$ varies continuously over the interval $[s_{\min},\,s_{\max}]$, the collection of pointwise confidence intervals defines a pointwise confidence envelope for the fitted functions.

\begin{rmk}
We compute confidence intervals both for the fitted bilognormal density as a whole and for each of its five parameters $\zeta,\,\mu_1,\,\sigma_1,\,\mu_2,\,\sigma_2$. For illustrative purposes, we report only the confidence interval for $\zeta$. Let $\boldsymbol{\zeta}\in\R^{2\times R}$ be the matrix collecting the values of $\zeta$ obtained from all simulations. Then, for each $a$,
$$ \IC_R^\delta(\boldsymbol{\zeta})=\left[\bar\zeta^a-t^{\delta/2}_{R-1}\sqrt{\frac{v^a}{R}},\,\bar\zeta^a+t^{\delta/2}_{R-1}\sqrt{\frac{v^a}{R}}\right]. $$
\label{ossBiLognormalParameterFitting}
\end{rmk}

\subsubsection{Empirical complementary cumulative distribution function}
\label{secEmpiricalCCDF}
The second and third rows of Figures~\ref{figSERStableReferenceConfiguration}--\ref{figSSERReferenceConfiguration} show the empirical Complementary Cumulative Distribution Function (CCDF) over the tail of the city-size distribution.

To characterise the tail of the distribution quantitatively, we take the upper quartile as threshold, i.e., the value $s_{3/4}$ of the city size $s$ such that the probability mass for $s\leq s_{3/4}$ is exactly $3/4$. More precisely, $\operatorname{Prob}(S\leq s_{3/4})=3/4$. In practice, this amounts to identifying the element $s_{3/4}$ of the tensor $\mathbf s^{T,a}_{\cdot,r}$ such that
$$ \left\vert\left\{s^{T,a}_{i,r}\geq s_{3/4}\,:\, i=1,\,\dots,\,N\right\}\right\vert=\left\lceil\frac{N}{4}\right\rceil=:N_{1/4},
    \qquad \forall\,a,\,r, $$
so that at least one quarter of the entries of $\mathbf s^{T,a}_{\cdot,r}$ are greater than or equal to $s_{3/4}$. We then regard the values $s\in [s_{3/4},\,s_{\max}]$ as constituting the tail of the distribution.

The empirical CCDF, denoted by $\CCDF_{\mathcal E}(\cdot;\,\mathbf s^{T,a}_{\cdot,r}):\R_+\to\R_+$, provides an empirical estimate of the probability $\operatorname{Prob}(S>y)$ and is defined by
\begin{equation}
    \CCDF_{\mathcal E}(y;\,\mathbf s^{T,a}_{\cdot,r}):=1-\frac{1}{N}\sum_{i=1}^N\chi_{[0,\,y]}(s^{T,a}_{i,r}),
    \label{eqComplementaryCumulativeDistributionFunction}
\end{equation}
where $\chi_{[0,\,y]}$ denotes the characteristic function of the interval $[0,\,y]$. Since~\eqref{eqComplementaryCumulativeDistributionFunction} is evaluated only for $y\in [s_{3/4},\,s_{\max}]$, we introduce the tensor $\mathbf p\in\mathbb R^{2\times N_{1/4}\times R}$ corresponding to the restriction of \(\mathbf s\) to its tail:
$$ p^a_{i,r}:=s^{T,a}_{N-N_{1/4}+i,r}, \qquad i=1,\,2,\,\dots,\,N_{1/4}. $$
It then follows that
\begin{align*}
    \CCDF_{\mathcal E}(y;\,\mathbf s^{T,a}_{\cdot,r}) &= 1-\frac{N-N_{1/4}}{N}-\frac{1}{N}\sum_{i=1}^{N_{1/4}}\chi_{[0,\,y]}(s^{T,a}_{N-N_{1/4}+i,r}) \\
    &= \frac{N_{1/4}}{N}\left[1-\frac{1}{N_{1/4}}\sum_{i=1}^{N_{1/4}}\chi_{[0,\,y]}(p^{T,a}_{i,r})\right] \\
    &= \frac{N_{1/4}}{N}\CCDF_{\mathcal E}(y;\,\mathbf p^a_{\cdot,r})
\end{align*}
for all $a$ and $r$. Thus, $\CCDF_{\mathcal E}(y;\,\mathbf p^a_{\cdot,r})$ and $\CCDF_{\mathcal E}(y;\,\mathbf s^{T,a}_{\cdot,r})$ differ only by the scaling factor $\frac{N_{1/4}}{N}$.
                
\begin{rmk}
Suppose that the city-size distribution exhibits a Pareto tail. Then, by~\eqref{eqComplementaryCumulativePareto}, $\CCDF(s)\approx\frac{c}{s^\beta}$, which, on a logarithmic scale, becomes $\log{\CCDF(s)}\approx-\beta\log{s}+\log{c}$. On a log--log plot, this corresponds to a straight line with slope $-\beta$ and intercept $\log{c}$. Now let $\tilde{\CCDF}:=k\CCDF$ be a rescaling of $\CCDF$ by a factor $k$. Then $\log{\tilde{\CCDF}(s)}=\log{\CCDF(s)}+\log{k}\approx-\beta\log{s}+\log{(kc)}$, which has the same slope as the original straight line. Hence, the Pareto index of the tail is invariant under multiplicative rescaling of the CCDF.
\label{ossLogarithmParetoCCDF}
\end{rmk}

In view of Remark~\ref{ossLogarithmParetoCCDF}, we may therefore plot $\CCDF_{\mathcal E}(y;\,\mathbf p^a_{\cdot,r})$ on a logarithmic scale to improve readability, since our primary interest does lies not in the pointwise values of $\CCDF$, but in the parameter $\beta$. This, however, gives rise to a minor issue: the last element $p^a_{N_{1/4},r}$ has zero ordinate and is therefore not visible on a logarithmic scale. To circumvent this issue, for graphical purposes we shift the empirical CCDF on the linear scale by setting
$$ \CCDF_{\mathcal E}(y;\,\mathbf p^a_{\cdot,r})+\frac{1}{2N_{1/4}} $$
for all $a$ and $r$. On a logarithmic scale, this transformation results in a non-uniform shift of the values, which are therefore slightly distorted. The distortion is small, however, since $1/(2N_{1/4})$ is small. This transformation is thus used solely for graphical purposes and not in the computation of the Pareto index $\beta$, which is always estimated directly from the vector $\mathbf p^a_{\cdot,r}$.

Overall, in the second and third rows of Figures~\ref{figSERStableReferenceConfiguration}--\ref{figSSERReferenceConfiguration} the empirical CCDF is plotted on a logarithmic scale using the normalised scatter plot of the points
$$ \left(\frac{N}{N_{1/4}}\CCDF_{\mathcal E}(\bar{p}^a_i;\,\mathbf s^{T,a}_{\cdot,r})+\frac{1}{2N_{1/4}},\,\bar{p}^a_i\right)
    =\left(\CCDF_{\mathcal E}(\bar{p}^a_i;\,\mathbf p^a_{\cdot,r})+\frac{1}{2N_{1/4}},\,\bar{p}^a_i\right) $$
for all $a$ and $i$, where $\bar{p}^a_i$ is the sample mean associated with the confidence interval
$$  \IC_R^{\delta}(\mathbf p^a_i)=\left[\bar{p}^a_i-t^{\delta/2}_{R-1}\sqrt{\frac{v^a_i}{R}},\,\bar{p}^a_i+t^{\delta/2}_{R-1}\sqrt{\frac{v^a_i}{R}}\right]. $$

\subsubsection{Pareto CCDF and Pareto index}
\label{secParetoCCDF}
The second and third rows of Figures~\ref{figSERStableReferenceConfiguration}--\ref{figSSERReferenceConfiguration} also show the Pareto CCDF fitted over the tail of the city-size distribution.

To fit the Pareto distribution, we proceed analogously to Section~\ref{secEmpiricalCCDF}. Let $\CCDF_{\mathcal P}(\cdot;\,\mathbf p^a_{\cdot,r}):\R_+\to\R_+$ denote the CCDF of the Pareto distribution, cf.~\eqref{eqComplementaryCumulativePareto}, fitted to the vector $\mathbf p^a_{\cdot,r}$, and let $\CCDF_{\mathcal P}(\cdot;\,\mathbf p^a):\R_+\to\R_+$ be the vector-valued function defined by
$$ \CCDF_{\mathcal P}(y;\,\mathbf p^a):=
    \left[\CCDF_{\mathcal P}(y;\mathbf p^a_{\cdot,1}),\,\CCDF_{\mathcal P}(y;\,\mathbf p^a_{\cdot,2}),\,\dots,\,\CCDF_{\mathcal P}(y;\,\mathbf p^a_{\cdot,R})\right]^\top. $$
The corresponding confidence intervals are then given by
$$ \IC_R^{\delta}\!\left(\frac{N}{N_{1/4}}\CCDF_{\mathcal P}(y;\,\mathbf p^a)\right)
    =\left[\bar{\CCDF}_{\mathcal P}^a(y)-t^{\delta/2}_{R-1}\sqrt{\frac{v^a(y)}{R}},\,\bar{\CCDF}_{\mathcal P}^a(y)+t^{\delta/2}_{R-1}\sqrt{\frac{v^a(y)}{R}}\right],
    \qquad y\in [s_{3/4},\,s_{\max}], $$
where $N/N_{1/4}$ is the normalising scaling factor introduced in the previous section, which is required here to compare the fitted Pareto CCDF with the empirical CCDF.

\begin{rmk}
As in Remark~\ref{ossBiLognormalParameterFitting}, confidence intervals are computed both for the fitted Pareto CCDF and for its parameter $\beta$. In particular, if $\boldsymbol\beta\in\R^{2\times R}$ denotes the matrix collecting the Pareto indices obtained from all simulations, then
$$ \IC_R^\delta(\boldsymbol\beta)=\left[\bar\beta^a-t^{\delta/2}_{R-1}\sqrt{\frac{v^a}{R}},\,\bar\beta^a+t^{\delta/2}_{R-1}\sqrt{\frac{v^a}{R}}\right] $$
for each $a$.
\end{rmk}
        
\subsubsection{Bimodal lognormal CCDF}
The third rows of Figures~\ref{figSERStableReferenceConfiguration}--\ref{figSSERReferenceConfiguration} also show the bilognormal CCDF fitted over the tail of the city-size distribution.

Let $\CCDF_{\mathcal L}(\cdot;\,\mathbf p^a_{\cdot,r}):\R_+\to\R_+$ and $\CCDF_{\mathcal B}(\cdot;\mathbf p^a_{\cdot,r}):\R_+\to\R_+$ denote the CCDFs of the lognormal and bilognormal distributions, respectively, fitted to the vector $\mathbf p^a_{\cdot,r}$. By linearity, it follows from~\eqref{eqBimodalLognormalDistribution} that
$$ \CCDF_{\mathcal B}(y;\,\mathbf p^a_{\cdot,r})=\zeta\CCDF_{\mathcal L}(y;\,\mathbf p^a_{\cdot,r})
    +(1-\zeta)\CCDF_{\mathcal L}(y;\,\mathbf p^a_{\cdot,r}). $$
Hence, analogously to Section~\ref{secParetoCCDF}, we can define the vector-valued function $\CCDF_{\mathcal B}(\cdot;\,\mathbf p^a):\R_+\to\R_+$ by
$$ \CCDF_{\mathcal B}(y;\,\mathbf p^a):=
    \left[\CCDF_{\mathcal B}(y;\,\mathbf p^a_{\cdot,1}),\,\CCDF_{\mathcal B}(y;\,\mathbf p^a_{\cdot,2}),\,\dots,\,\CCDF_{\mathcal B}(y;\,\mathbf p^a_{\cdot,R})\right]^\top. $$
The corresponding confidence intervals are then given by
$$ \IC_R^{\delta}\!\left(\frac{N}{N_{1/4}}\CCDF_{\mathcal B}(y;\,\mathbf p^a)\right)
    =\left[\bar{\CCDF}_{\mathcal B}^a(y)-t^{\delta/2}_{R-1}\sqrt{\frac{v^a(y)}{R}},\,
        \bar{\CCDF}_{\mathcal B}^a(y)+t^{\delta/2}_{R-1}\sqrt{\frac{v^a(y)}{R}}\right],
            \qquad y\in [s_{3/4},\,s_{\max}] $$
for each $a$.
        
\subsubsection{Data transformation}
The plot in the second to fourth rows of Figures~\ref{figSERStableReferenceConfiguration}--\ref{figSSERReferenceConfiguration} are shown on a semilogarithmic scale. This choice is motivated by the fact that a lognormally distributed dataset $\mathbf s$ is transformed into a normally distributed dataset $\mathbf h\in\R_+^{N_f\times2\times N\times R}$ by the transformation
\begin{equation}
    h_{i,r}^{n,a}:=\log_{10}{s_{i,r}^{n,a}},
    \label{eqlog10transformation}
\end{equation}    
which facilitates interpretation, particularly when the parameters are varied, as in parametric studies presented in the Supplementary Material.

Under the transformation~\eqref{eqlog10transformation}, the histogram bins are obtained as $\mathbf b^a_{\cdot,r}:=\mathcal{B}(\mathbf h^{T,a}_{\cdot,r})$, and the confidence intervals are transformed accordingly. Moreover, the probability density $\bar f_N(s):=\bar f_N(s,T)$ is mapped to a new density, denoted by $\bar l_N=\bar l_N(s)$, while preserving the probability mass:
$$ \bar f_N(s)\,ds=\bar l_N(h)\,dh. $$
Since $h=\log_{10}{s}$, it follows that
$$ \bar l_N(h)=10^h\log{10}\bar f_N(10^h). $$

Note that~\eqref{eqLognormalDistribution} uses the natural logarithm, whereas the transformation~\eqref{eqlog10transformation} is defined in terms of the base-$10$ logarithm. The two logarithmic scales differ by a factor of $\log{10}\approx2.303$, which is precisely the Jacobian factor appearing in the expression for $\bar l_N(h)$ above. Accordingly, expressing the lognormal distribution~\eqref{eqLognormalDistribution} in terms of base-$10$ logarithms requires rescaling its parameters as $\mu\mapsto\mu/\log{10}$ and $\sigma\mapsto\sigma/\log{10}$. Under this reparametrisation, the prefactor in~\eqref{eqLognormalDistribution} becomes $\left(x\sigma\sqrt{2\pi}\,\log{10}\right)^{-1}$. Thus, the parameters $\mu_{10}:=\mu/\log{10}$ and $\sigma_{10}:=\sigma/\log{10}$ obtained from the fits reported in this work are expressed on a base-$10$ logarithmic scale and cannot be directly compared with the corresponding natural-logarithm parameters in~\eqref{eqLognormalDistribution}. The first rows of Figures~\ref{figSERStableReferenceConfiguration}--\ref{figSSERReferenceConfiguration} therefore display the transformed density $\bar l_N$.

By contrast, the empirical CCDFs remain unchanged in value, with their domain simply mapped from $[s_{\min},\,s_{\max}]$ to $[\log_{10}{s_{\min}},\,\log_{10}{s_{\max}}]$. This follows from the fact that the logarithm is strictly increasing and therefore preserves the ordering of the data. In particular:
$$ \operatorname{Prob}(S<s)=\operatorname{Prob}(\log_{10}{S}<\log_{10}{s}), \qquad \forall\,s\in\R_+. $$

Finally, we observe that, on a semilogarithmic scale, the relationship discussed in Remark~\ref{ossLogarithmParetoCCDF} takes the form $\log_{10}{\tilde\CCDF(h)}\approx-\beta h+\log_{10}{(kc)}$, which remains linear in $h$.

\subsubsection{Mean city size against the city degree}
The fourth rows of Figures~\ref{figSERStableReferenceConfiguration}--\ref{figSSERReferenceConfiguration} show scatter plots of the mean city size against the city degree.

We compute the mean city size by averaging over the ensemble of simulations:
$$ \bar s^a_i:=\frac{1}{R}\sum_{r=1}^Rs^{T,a}_{i,r} $$
for all $a$ and $i$, and then plot the set of points $\{(\xi_i,\,\bar s^a_i)\}_{a,\,i}$.

In this case, we do not show the corresponding confidence intervals, as they would make the plots, which are already rather dense, difficult to read.
        
\subsubsection{Evolutions of the mean city size by degree class}
\label{secEvolutionMeanCitySizeByDegreeClass}
These are the plots shown in the fifth rows of Figures~\ref{figSERStableReferenceConfiguration}--\ref{figSSERReferenceConfiguration}.

Let $\mathcal X\subseteq\{1,\,2,\,\dots,\,N\}$ denote the set of distinct degrees in the graph, and let us retain the notation $\bar{\mathbf s}\in\R^{N_f\times 2\times\vert\mathcal X\vert}$ for the tensor whose entries are defined by
$$ \bar s^{n,a}_k:=\frac1{\vert\mathcal I_k\vert}\sum_{i\in\mathcal I_k}\frac{1}{R}\sum_{r=1}^Rs^{n,a}_{i,r} $$
for all $n$, $a$, and $k$, where $\mathcal I_k:=\{i\in\mathcal I\,:\,x_i=k\}\subseteq\mathcal I$ denotes the set of indices of the vertices of degree $k$.

By linearly interpolating the points $\{(t^f_n,\,\bar s^{n,a}_k)\}_{n,\,a,\,k}$ between consecutive time instants, we represent the time evolution of the mean population size of cities associated with vertices of degree $k$, averaged over all simulations. We use this plot to verify that the system has reached a steady state by the prescribed final simulation time $T$.
               
As in the previous section, we do not show the corresponding confidence intervals, for the same reason given there.

\section*{CRediT authorship contribution statement}
Valerio Taralli: Conceptualization, Methodology, Software, Formal analysis, Investigation, Writing - Original Draft. Andrea Tosin: Supervision.

\bibliographystyle{cas-model2-names}
\bibliography{biblio}

\includepdf[pages=-]{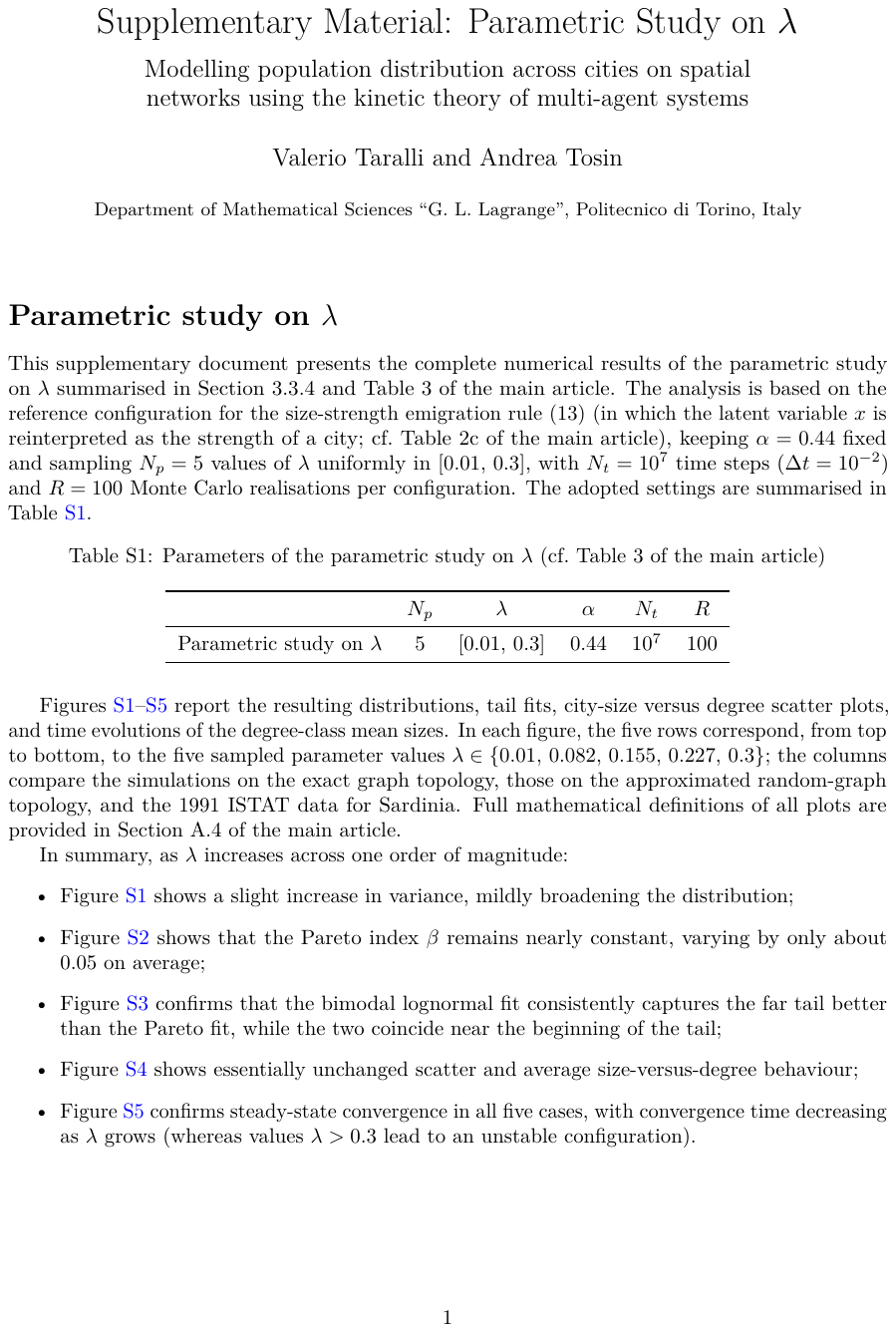}
\includepdf[pages=-]{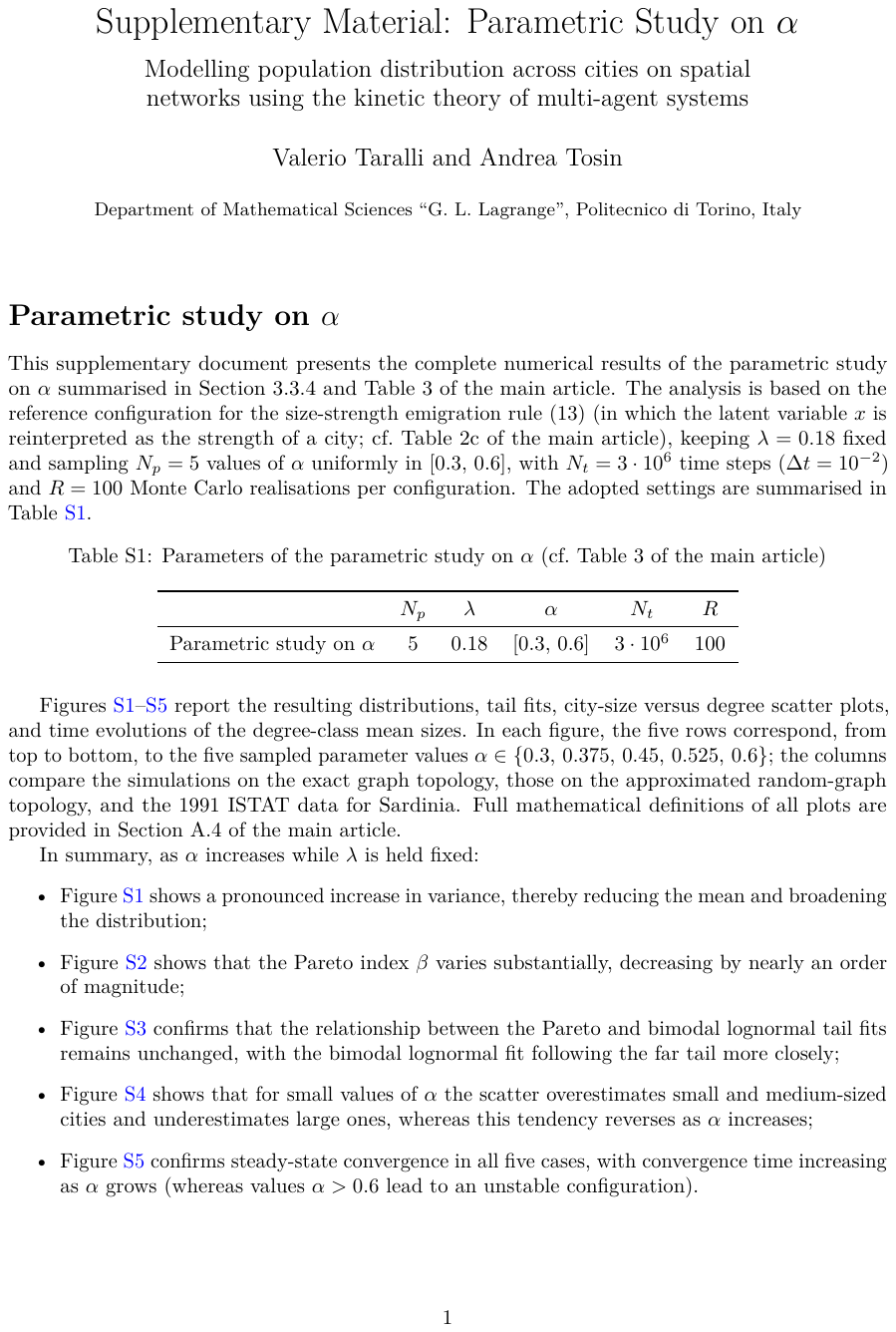}

\end{document}